\documentclass[a4paper,11pt]{article}
\usepackage{etex}
\usepackage[T1]{fontenc}
\usepackage[utf8]{inputenc}

\usepackage{geometry}
\usepackage{xcolor}
\usepackage{amsmath, array, amssymb, amsfonts}
\usepackage{mathtools}
\usepackage{amsthm}
\usepackage{dsfont}

\usepackage[french, english]{babel}
\usepackage[all]{xy}

\usepackage{fourier-orns}
\usepackage{textcomp}
\usepackage{lmodern}

\usepackage{array,booktabs}
\usepackage{caption}
\usepackage[colorlinks=true, pdfstartview=FitV, linkcolor=blue, citecolor=blue, urlcolor=blue]{hyperref}
\usepackage[square,numbers,sort,compress,semicolon,merge]{natbib}
\usepackage{notoccite}

\usepackage{empheq}
\usepackage{cancel}
\usepackage{tikz}
\usepackage{tikz-cd}
\usetikzlibrary{matrix,arrows,external}

\let\ssection=\section
\renewcommand{\section}{\setcounter{equation}{0}\ssection}

\DeclareMathAlphabet{\mathpzc}{OT1}{pzc}{m}{it}

\newcommand{\bs}[1]{\boldsymbol{#1}}
\newcommand{\defeq}{\vcentcolon=}
\newcommand{\rdefeq}{=\vcentcolon}
\renewcommand\P{\mathcal{P}}
\newcommand\M{\mathcal{M}}

\newcommand\doubleR{\mathbb{R}}

\renewcommand\1{\textbf{1}}

\renewcommand\H{\mathcal{H}}

\renewcommand\L{\mathcal{L}}

\newcommand\U{\mathcal{U}}
\newcommand\SO{\mathcal{SO}}

\newcommand\K{\mathcal{K}}

\newcommand\rarrow{\rightarrow}
\newcommand\xrarrow{\xrightarrow}

\newcommand\LieG{\mathfrak{g}}
\newcommand\LieH{\mathfrak{h}}

\newcommand\co{\mathfrak{co}}

\renewcommand\tilde{\widetilde}
\newcommand\h{\widehat}

\newcommand\w{\wedge}
\renewcommand\d{\partial}

\renewcommand\-{^{-1}}

\renewcommand\1{\mathds{1}}

\DeclareMathOperator{\Id}{Id}
\DeclareMathOperator{\Diff}{Diff}
\DeclareMathOperator{\Aut}{Aut}

\newcommand\s{\sigma}

\theoremstyle{definition}

\hypersetup{
pdfauthor={J. François, S. Lazzarini, T. Masson},
pdftitle={BRST structure for the mixed Weyl--diffeomorphism residual symmetry},
pdfsubject={},
pdfcreator={pdflatex},
pdfproducer={pdflatex},
pdfkeywords={}
}

\date{}
\begin{document}

\begin{titlepage}
\title{BRST structure for the mixed Weyl--diffeomorphism residual symmetry\\[2mm]}
\author{J. François,${\,}^a$\footnote{Supported by a Riemann fellowship.} \hskip 1.5mm
 S. Lazzarini${\,}^b$ and T. Masson${\,}^b$}
\date{Published version}
\maketitle
\begin{center}

\vskip -5mm
${}^a$ Riemann Center for Geometry and Physics,\\
Leibniz Universität Hannover,\\
Appelstr. 2, 30167 Hannover, Germany
\\[3mm]
${}^b$ Centre de Physique Théorique,\\
Aix Marseille Université \& Université de Toulon \&  CNRS UMR 7332,\\
13288 Marseille, France
\end{center}

\maketitle

\vskip 1cm
\hfill{\em To the memory of Daniel Kastler (1926-2015)}

\vskip 2cm

\begin{abstract}
In this paper, we show the compatibility of the so-called ``dressing field method'', which allows a systematic reduction of gauge symmetries, with the inclusion of diffeomorphisms in the BRST algebra of a gauge theory. 
The robustness of the scheme is illustrated on two examples where Cartan connections play a significant role. The former is General Relativity, while the latter concerns the second-order conformal structure where one ends up with a BRST algebra handling both the Weyl residual symmetry and diffeomorphisms of spacetime. We thereby provide a geometric
counterpart to the BRST cohomological treatment used in~\cite{Boulanger1} in the construction of a Weyl covariant tensor calculus.

\end{abstract}


\end{titlepage}

\newpage


\tableofcontents

\vspace{1cm}

\section{Introduction}  
\label{Introduction}  

Modern Field Theory framework (classical and quantum), to this day so successful in describing Nature from particles to cosmology, rests on few keystones, one of which being the notion of local symmetry. Elementary fields are subject to local transformations which are required to leave invariant the physical theory (the Lagrangian). These transformations thus form a symmetry of the theory.  Requiring local symmetries is such a stringent restriction on the admissible theories and their content so as to justify Yang's well known aphorism: ``symmetry dictates interaction'' \cite{YangSelPap}.

\medskip
Confirmed fundamental theories distinguish two types of symmetries; ``external'' symmetries stemming from transformations of spacetime $\M$, that is diffeomorphisms $\Diff(\M)$, and ``internal'' symmetries stemming from the action of a gauge group $\H$. 

From a geometric standpoint, $\H$ is (isomorphic to) the group of vertical automorphisms $\Aut_V(\P)$ of a principal fiber bundle $\P(\M, H)$ over spacetime $\M$ with structure group $H$, itself a subgroup of the group of bundle automorphisms $\Aut(\P)$. While $\Aut_V(\P)$ projects onto identity map of $\M$, $\Aut(\P)$ projects onto $\Diff(\M)$. The group $\Aut(\P)$ offers a geometrical way to gather both internal and spacetime  symmetries through the short exact sequence, 
\begin{align*}
\lbrace I \rbrace \rarrow \H \rarrow \Aut(\P) \rarrow \Diff(\M)\rarrow \lbrace \Id_\M \rbrace.
\end{align*}

It is often easier to work with the infinitesimal version of the transformations, that is with the Lie algebras of the symmetry groups.
As a matter of fact, the infinitesimal gauge transformations are encoded in the so-called BRST differential algebra of a gauge theory in which the infinitesimal local gauge parameter is turned into the Faddeev-Popov ghost field. This is algebraic in nature~\cite{Sto84}. The Lie algebra of $\Diff(\M)$ is the space of smooth vector fields $\Gamma(T\M)$ on $\M$ with the Lie bracket of vector fields. This is geometric in nature. The corresponding infinitesimal symmetries are summed up in the following short exact sequence of Lie algebroids, 
\begin{align*}
0 \rarrow \text{Lie}\,\H \rarrow \Gamma_H(\P) \rarrow \Gamma(T\M) \rarrow 0 \ ,
\end{align*}
where $\Gamma_H(\P):=\text{Lie}\Aut(\P)$ are the $H$-right-invariant vector fields on $\P$. Since both infinitesimal symmetries (internal/external or algebraic/geometric) ought to be unified in the central piece of the above sequence, one expects to find a BRST treatment that encompasses both (pure) gauge transformations and diffeomorphisms.

This problem has been already addressed by several authors. Pioneering work is~\cite{Baulieu:1984pf}, and improved in~\cite{Baulieu:1985md}. A refined work addressing the case of pure gravity is~\cite{Langouche-Schucker-Stora}. 
From these papers, a general heuristic construction emerges that
allows to alter a pure gauge BRST algebra so as to obtain a
\emph{shifted}~\footnote{We use the denomination introduced in~\cite{Bertlmann}.} BRST algebra that describes both together gauge and diffeomorphism symmetries.
Roughly, this shifting operation amounts to introducing the
diffeomorphism ghost (vector field) $\xi$ and modifying accordingly the ``Russian formula'' (or ``horizontality condition'') and the BRST operator~$s$ itself.

\medskip
In a previous work~\cite{GaugeInv} we proposed a systematic approach to reduce gauge symmetries by the \emph{dressing field method}. Its relevance to recent controversies on the proton spin decomposition was advocated in \cite{NucleonSpin}, and its generalization to higher-order $G$-structures was suggested in \cite{FLM15} by application to the second-order conformal structure (see~\cite{JF_PhD} for the general case). In the latter, it was also shown that the dressing field method adapts to the BRST framework: from an initial pure gauge BRST algebra one obtains, by dressing, a \emph{reduced} BRST algebra describing residual gauge transformations and whose central object is the \emph{composite ghost} which encapsulates the residual gauge symmetry (if any). 

\medskip
The aim of the present paper is to combine together the shifting operation and the dressing field method providing a \emph{residual shifted} BRST algebra that describes both residual gauge transformations and diffeomorphisms, for which the central object is the \emph{dressed shifted ghost}.
In doing so, we address the issue of their compatibility and we provide the necessary condition for the two operations of shifting and dressing to commute between themselves. A pragmatic criterion for the failure of that condition is discussed. We then illustrate the construction on two examples: we shall first treat General Relativity. Then, notably enough we shall deal with  the second-order conformal structure where our scheme easily provides the  BRST structure of the residual mixed Weyl + diffeomorphism symmetry out of the whole conformal + diffeomorphism symmetry.

\medskip
The paper, which can be considered as a sequel of~\cite{FLM15}, is organized as follows. In section~\ref{Mixed BRST symmetry gauge + diffeomorphisms: the general scheme} we first recall the minimal definition of  the standard BRST approach, and then give the heuristic construction allowing to include diffeomorphisms of spacetime $\M$. In section~\ref{The dressing field method and diffeomorphisms} we provide the basics of the dressing field method, exhibit the reduced BRST algebra and the associated composite ghost, and finally show the compatibility with the inclusion of diffeomorphisms. We also exhibit the necessary and sufficient condition securing the commutation of the shifting and dressing operations. Section~\ref{Example: the geometry of General Relativity} deals with the simple application to General Relativity (GR). Section~\ref{Example: the second-order conformal structure} details the rich example of the second-order conformal structure. Finally, we discuss our results and conclude in section~\ref{Conclusion}.

\section{Mixed BRST symmetry gauge + Diff: a general scheme}   
\label{Mixed BRST symmetry gauge + diffeomorphisms: the general scheme}   

As just mentioned in the Introduction, the search for a single BRST algebra for the description of both gauge symmetries and diffeomorphisms has been quite early addressed. To the best of our knowledge, a pioneering work is~\cite{Baulieu:1984pf}. There, a first step was the recognition of the necessity to modify the so-called \emph{horizontality condition}~\cite{Ne'eman:1978gg,Baulieu:1981sb}, also named ``Russian formula''~\cite{Sto84}, encapsulating the standard BRST algebra. 

Then~\cite{Baulieu:1985md} significantly improved the previous work by generalizing it (to a wide class of supersymmetric Einstein-Yang-Mills theories), but first and foremost --besides modifying the horizontality condition-- the ghost field was modified as well. 
This change of generators has also been performed in~\cite{Langouche-Schucker-Stora,Sto84,Stora:2005tp} in the pure gravitational case, where a BRST algebra for a Lorentz $+$ diff symmetry (or {\em mixed} symmetry) in presence of a background field was given. 
\footnote{Global aspects require careful consideration which might lead one to look for a better adapted geometrical framework. }
For further references on the subject, see {\em e.g.}~\cite{Ban88,Bertlmann,Barnich:2000zw}. 

In this section however, we aim at giving the simplest heuristic construction allowing to modify the BRST algebra of a gauge (Yang-Mills) theory so as to include diffeomorphisms. Let us start by recalling the definition of a standard BRST gauge algebra.

\subsection{The BRST gauge algebra} 
\label{The BRST gauge algebra} 

The geometrical framework of Yang-Mills theories is that of a principal bundle $\P=\P(\M,H)$ over an $m$-dimensional spacetime $\M$, with structure group~$H$ whose Lie algebra is $\LieH$. Let $\omega \in \Omega^1(\P,\LieH)$ be a (principal) connection $1$-form on $\P$ and let $d\omega+ \tfrac{1}{2} [\omega, \omega]=:\Omega \in \Omega^2(\P, \LieH)$ be its curvature; 
let $\Psi$ denote a section of an associated bundle constructed out of a representation $(V, \rho)$ of~$H$.

In order to stick to the usual local description on an open set $\U\subset \M$ (through a local trivializing section of the principal bundle $\P$) 
the local connection 1-form gives the usual Yang-Mills gauge potential $A$ with field strength $F = dA+ \tfrac{1}{2} [A,A]$, and matter field $\psi: \U \rarrow V$.

To the infinitesimal generators of gauge transformations is associated a Faddeev-Popov ghost field $v: \U \rarrow \LieH^*\otimes\LieH$, where $\LieH^*$ is the dual Lie algebra $\LieH$ of $H$. 

The BRST algebra of a non-abelian gauge field theory is well-known~\cite{BRS-76} to be defined as
\begin{align}
\label{BRST_YM}
sA = -Dv := -dv -[A, v], \qquad  sF=[F, v],  \qquad s\psi=-\rho_*(v)\psi, \qquad  sv =-\tfrac{1}{2}[v, v]. 
\end{align}
Let us remind that the BRST operator $s$ is an antiderivation which anticommutes with the exterior differential $d$ and with odd differential forms, and $[\, ,\,]$ is a graded bracket with respect to the form+ghost degrees. It is easily verified that $s^2=0$. We shall denote by $BRST$ the above differential algebra.
\bigskip

This differential algebra can be incorporated into a larger differential algebra bigraded by the form and ghost degrees, whose nilpotent operator is $\tilde d:=d+s$ such that $\tilde d{\,}^2=0$. 
Accordingly, one may define the ``algebraic connection''~\cite{DbV86} $\tilde A:=A+v$ of bidegree $1$. 
Then, due to the definition of $s$ for the pure gauge sector of the differential algebra~\eqref{BRST_YM} one has the ``Russian formula''
\begin{align}
\label{RussForm}
\tilde d \,\tilde A + \tfrac{1}{2}[\tilde A, \tilde A] =  F \ .
\end{align}
Indeed, by expanding \eqref{RussForm} with respect to the ghost degree, one recovers the pure gauge sector of~\eqref{BRST_YM}.

In the same way, $\psi$ being a $0$-form stands alone in the bigraded algebra $\tilde \psi=\psi$, and if one requires the following horizontality condition~\cite{Baulieu:1985md},
\begin{align}
\label{HorCond}
\tilde D \tilde\psi :=\tilde d\, \tilde\psi + \rho_*(\tilde A)\tilde\psi =  D \psi,
\end{align}
one recovers the BRST variation of the matter sector in~\eqref{BRST_YM}.

These horizontality conditions \eqref{RussForm} and \eqref{HorCond}
provide a very convenient starting point that allows a systematic and straightforward inclusion of diffeomorphisms in the BRST framework.

\subsection{Adding diffeomorphisms} 
\label{Adding diffeomorphisms} 

The infinitesimal generators of diffeomorphisms are vector fields. According to the usual BRST setting, let us associate a ghost vector field $\xi$ to the infinitesimal diffeomorphism symmetry. Denote by $i_\xi$ its usual inner product on differential forms. The inner product is of degree $-1$ but $\xi$ has ghost number $1$, then $i_\xi$ is of total degree $0$ and is thus a derivation. The Lie derivative acting on differential forms is accordingly an antiderivation of degree $+1$ (graded Cartan formula) and yields the Cartan operation
\begin{align}
\label{Lie_der}
L_\xi &:= i_\xi\, d - d\, i_\xi\ , &  \text{with} \qquad [L_\xi, i_\xi]
&= i_{[\xi, \xi]} & \text{and} \qquad d\, L_\xi + L_\xi \, d = 0.
\end{align}
These identities will be extensively used in the sequel.

\smallskip
The BRST gauge algebra~\eqref{BRST_YM} is equivalently recast into the horizontality conditions \eqref{RussForm} and \eqref{HorCond}. To obtain a new BRST algebra that also takes diffeomorphisms into account, one may accordingly modify these horizontality conditions. A systematic way to do so rests on the following ansatz~\cite{Baulieu:1985md,Stora_private_comm}:
the new BRST operator $\sigma$ is defined through the intertwining
\begin{align}
\label{sigma}
d+\s \defeq e^{i_\xi}\,\tilde d \, e^{-i_\xi}
\end{align}
where $e^{i_\xi}=1+i_\xi +\tfrac{1}{2}i_\xi i_\xi+ \cdots$ is the formal power series of the exponential of $i_\xi$ and is shown \cite{Baulieu:1985md} to be a morphism of the exterior algebra of
differential forms and also a Lie algebra homomorphism, namely $e^{i_\xi}[\alpha, \beta] =[e^{i_\xi} \alpha, e^{i_\xi}\beta]$.

With the ansatz \eqref{sigma}, the Russian formula~\eqref{RussForm} becomes
\begin{align}
\label{shiftRussForm}
(d +\s)(e^{i_\xi} \tilde A) + \tfrac{1}{2}\left[e^{i_\xi}\tilde A, e^{i_\xi}\tilde A\ \right]= e^{i_\xi}F.
\end{align}
One thus readily computes for the algebraic connection 
\begin{align*}
e^{i_\xi}\tilde A = (1+i_\xi) (A+v) =: A +v + i_\xi A
\end{align*}
where, according to the ghost degree, we are led to define the \emph{shifted ghost}:
\begin{align}
\label{shiftghost}
v' \defeq v+i_\xi A.
\end{align}
In more detail, the Russian formula~\eqref{shiftRussForm} becomes:
\begin{align}
\label{RussForm'}
 (d+\s)(A +v') + \tfrac{1}{2}[A +v', A+v'] = F + i_\xi F + \tfrac{1}{2}i_\xi i_\xi F.
\end{align}
By sorting out the terms according to the bigrading one gets in a row:

\smallskip \noindent
Degree $(2, 0)$ corresponds to the usual Cartan structure equation, $dA +\tfrac{1}{2}[A,A]=F$. 

\smallskip \noindent
Degree $(1, 1)$ gives rise to
\begin{align}
\label{sig_om}
\s A=-Dv'+i_\xi F := - dv' - [A, v'] + i_\xi F.
\end{align}
From these two one easily finds $\s F = [F, v'] - D(i_\xi F)$. Degree $(0, 2)$ yields
\begin{align}
\label{sig_v'}
\s v' = - \tfrac{1}{2}[v', v'] + \tfrac{1}{2}i_\xi i_\xi F.
\end{align}

In the same way the horizontality condition for the matter fields reads,
\begin{align}
\label{hor_cond_psi}
(d+\s) e^{i_\xi}\psi + e^{i_\xi} \rho_*(A+v) \psi = e^{i_\xi}D\psi.
\end{align}
In ghost degree $1$ one finds, 
\begin{align}
\s \psi = - \rho_*(v') + i_\xi D\psi.
\end{align}
Moreover, requiring the nilpotency $\s^2=0$ on the generators $(A, \psi, v')$ leads to
\begin{align}
\label{sig_xi0}
\s \xi =  \tfrac{1}{2}[\xi, \xi]
\end{align}
where $[\xi, \xi]$ is the Lie bracket of vector fields.\footnote{As
  stated in~\cite{Mil84}, the Lie algebra of  diffeomorphisms is
  anti-isomorphic to the Lie algebra of vector fields. This explains
  why the factor $\tfrac{1}{2}$ occurs without a minus sign. Thus upon substituting $\xi$ by $-\xi$ one recovers variations obtained
  in~\cite{Langouche-Schucker-Stora}.}
Accordingly, the new shifted BRST algebra with generators $(A, \psi,  v', \xi)$ and $\s$-operation describing both infinitesimal transformations gauge + $\Diff(\M)$ is defined by
\begin{align}
\label{mixed_BRST}
\s A = -Dv' + i_\xi F&, \quad \s F = [F, v'] - D(i_\xi F), \quad \s\psi= - \rho_*(v') + i_\xi D\psi, \notag \\[2mm]
 \s v'&= -\tfrac{1}{2}[v', v'] +\tfrac{1}{2}i_\xi i_\xi F, \qquad  \s \xi =  \tfrac{1}{2}[\xi, \xi].
\end{align}

\medskip
It is somewhat hard to disentangle the two symmetries with the above presentation of the shifted algebra. But one can give an alternative  presentation which relies on the fact that the shifted ghost $v'$ assumes the form \eqref{shiftghost}. Having taken this into account, one can give the action of $\s$ on the generators $(A, \psi, v, \xi)$, and \eqref{mixed_BRST} becomes, 
\begin{align}
\label{mixed_BRST2}
\s A = -Dv+ L_\xi A&, \quad \s F=[F, v] + L_\xi F, \quad \s \psi =-\rho_*(v) \psi + L_\xi \psi, \notag \\[2mm]
\s v&=-\tfrac{1}{2}[v, v] +L_\xi v, \quad \s\xi= \tfrac{1}{2}[\xi, \xi].
\end{align}
This presentation shows that $\s =s + L_\xi$, so that the actions of
gauge and diffeomorphisms symmetries turn out to be decoupled on this
set of generators. For convenience and subsequent purpose, let us
denote $BRST^\xi$ either of the two presentations \eqref{mixed_BRST} or \eqref{mixed_BRST2} for the resulting shifted BRST algebra.

\section{The dressing field method and diffeomorphism symmetry} 
\label{The dressing field method and diffeomorphisms} 

A systematic approach to reduce gauge symmetries has been proposed and applied to various examples in \cite{GaugeInv, NucleonSpin, FLM15, JF_PhD} (see also \cite{Masson-Wallet}). It is already compatible with the BRST framework, as it will be briefly outlined in the following. 
It remains to study the compatibility with the shifting procedure described above. This will be the main issue of this section. 

\subsection{The dressing field method: a primer} 
\label{The dressing field method: a primer} 

The gauge group of a Yang-Mills theory  is defined as
$\H:=\left\{ \gamma :\U \rarrow H\right\}$ and it carries the canonical action on itself $\gamma_1^{\gamma_2}=\gamma_2\- \gamma_1 \gamma_2$ for any $\gamma_1, \gamma_2  \in \H$. It respectively acts on gauge potential, field strength and matter fields according to,
\begin{align}
\label{GT}
A^\gamma=\gamma\-A\gamma + \gamma\*d\gamma, \qquad F^\gamma=\gamma\-F\gamma, \qquad \text{and} \qquad \psi^\gamma =\rho(\gamma\-) \psi.
\end{align}
Suppose the theory also contains a (Lie) group-valued field
$u:\U\rarrow G'$ defined by its transformation under  $\H'=\left\{ \gamma':\U\rarrow H'\right\}$, where $H' \subseteq H$ is a subgroup, according to 
  \[
u^{\gamma'}:={\gamma'}\-u\ , \quad \gamma'\in\H'.
\]
One can define the following {\em composite fields} \footnote{This
  means that $G'$ is to be suitable for this definition, in particular
  it shares the same representations as $H$ (at least the adjoint representation and $\rho$).}
\begin{align}
\label{comp_fields}
\h A := u\- A u + u\-du,  \qquad \h F :=u\-F u \qquad \text{and} \qquad \h \psi:= \rho(u\-)\psi.
\end{align}
The Cartan structure equation still holds, $\h F=d\h A + \tfrac{1}{2}[\h A, \h A]$. 
\medskip

Despite the formal similarity with \eqref{GT}, the composite fields given in~\eqref{comp_fields} are not mere gauge transformations since $u\notin \H$, as is testified by its transformation property under $\H'$ and the fact that in general $G'$ may be different from $H$. This fact clearly implies also that the composite field $\h A$ does no longer belong to the space of local connections. 

Finally, as it can be easily checked, the composite fields \eqref{comp_fields} are
$\H'$-invariant and are only subject to residual gauge transformation
laws in $\H \setminus \H'$. \footnote{To some extent, $H'$ is identifed to be a subgroup of $H$ along which the gauge invariance can be restored. One may also check that the complement $\H \setminus \H'$ is stable under $H$.}
In the case where $H'=H$, these composite fields are $\H$-gauge invariants and may become good candidates to be observables. 
\medskip

It is easy to show that the BRST algebra pertaining to a pure gauge theory is modified by the dressing as
\begin{align}
\label{dressed_BRST}
s\h A =-\h D \h v=-d\h v - [\h A, \h v\,], \quad s\h F=[\h F, \h v\,], \quad s\h \psi=-\rho_*(\h v\,)\h\psi, \qquad s\h v =-\tfrac{1}{2}[\,\h v ,\h v\,],
\end{align}
upon defining the {\em composite ghost}
\begin{align}
\label{dressed_ghost}
\h v:= u\- v u + u\- s u \ .
\end{align}
To the best of our knowledge, first occurrences of such a change of
generator in a BRST setting for specific cases can be found
in~\cite{Garajeu-Grimm-Lazzarini} and~\cite{Lazzarini-Tidei}. Let us denote by $\h{\text{BRST}}$ the above dressed BRST algebra. The results~\eqref{dressed_BRST} and \eqref{dressed_ghost} are actually strictly formal. They do not depend on the fact that $u$ is a dressing field, namely on an explicit expression of the variation~$su$. See~\cite{FLM15} for a detailed discussion on this point. 
When $u$ is indeed a dressing field, 
\eqref{dressed_ghost} may thus encode the infinitesimal residual gauge symmetry, if any. If $\h v=0$, obviously the differential algebra $\h{\text{BRST}}$ becomes trivial, thus expressing the whole gauge invariance of the composite fields. 
\medskip

As in the usual case, upon defining the {\em composite algebraic connection}
\begin{align*}
\h A + \h v = u\- \tilde A u + u\- \tilde d u
\end{align*}
the dressed algebra \eqref{dressed_BRST} can be compactly encapsulated
into the following two horizontality conditions
\begin{align}
\label{dressed_HorCond}
(d+s) (\h A + \h v\,) + \tfrac{1}{2}[\h A + \h v, \h A + \h v\,] &= \h F, \qquad \text{and} \qquad (d+s) \h \psi + \rho_*(\h A +\h v\,) \h\psi = \h D \h \psi.
\end{align}

\subsection{Shifting and dressing} 
\label{Shifting and dressing} 

We now investigate the compatibility of the two operations of shifting (adding diffeomorphisms) and dressing. Two approaches are available to us.
\medskip

First, one can proceed as for the dressing of the initial BRST algebra
in order to obtain the algebra $\h{\text{BRST}}$. This amounts to expressing the initial gauge variables $(A, F, \psi)$ as functions of the dressed variables $(\h A, \h F, \h \psi)$ and the dressing field $u$, and replacing them into the \emph{first} presentation \eqref{mixed_BRST} of $BRST^\xi$. One then obtains, 
\begin{align}
\label{dressed_shifted_BRST}
\s \h A = -\h D \h{v'} + i_\xi \h F&,\qquad \s \h F= [\h F, \h{v'}] - \h D(i_\xi\h F), \qquad \s \h \psi=-\rho_*(\h{v'})\h\psi + i_\xi\h D\h\psi,\notag \\[2mm]
 \s \h{v'}&= -\tfrac{1}{2}[ \h{v'}, \h{v'}] +\tfrac{1}{2}i_\xi i_\xi \h F, \qquad  \s \xi =  \tfrac{1}{2}[\xi, \xi],
\end{align}
with the composite shifted ghost defined by
\begin{align}
\label{dressed_shifted_ghost}
\h{v'} := u\- v' u + u\-\s u.
\end{align}
Let us denote by $\h{BRST^\xi}$ this algebra. Since it assumes the
same formal presentation as \eqref{mixed_BRST}, one verifies that
$\s^2=0$ on $(\h A,  \h \psi, \h{v'})$ implies $\s \xi =
\tfrac{1}{2}[\xi, \xi]$ in the same
way.\footnote{\label{Second_Presentation} Likewise, performing the
  same substitution starting from the
  \emph{second} presentation \eqref{mixed_BRST2} of BRST$^\xi$ would
  result in an algebra for the composite fields formally identical to
  \eqref{mixed_BRST2}, still denoted by $\h{\text{BRST}^\xi}$, but with
  pure gauge ghost $(u^{-1}vu+u^{-1}\s u) - u^{-1}L_\xi u$.}
 
\medskip
The second possible route amounts to modifying the dressed horizontality conditions~\eqref{dressed_HorCond} by using the ansatz~\eqref{sigma},
\begin{align}
&(d +\s)e^{i_\xi} (\h A + \h v) + \tfrac{1}{2}[e^{i_\xi}(\h A+\h v), e^{i_\xi}(\h A+\h v)]= e^{i_\xi}\h F, \notag \\
&(d+\s) e^{i_\xi}\h\psi + e^{i_\xi} \rho_*(\h A+\h v) \psi = e^{i_\xi}\h D\h \psi.
\end{align}
Expansion according to the ghost degree provides, besides the Cartan structure equation for~$\h F$, 
\begin{align}
\label{shifted_dressed_BRST}
\s \h A = -\h D \h{v}' + i_\xi \h F&,\qquad \s \h F= [\h F, \h{v}'] - \h D(i_\xi\h F), \qquad \s \h \psi=-\rho_*(\h{v}')\h\psi + i_\xi\h D\h\psi,\notag \\[2mm]
 \s \h{v}'&= -\tfrac{1}{2}[ \h{v}', \h{v}'] +\tfrac{1}{2}i_\xi i_\xi \h F, \qquad  \s \xi =  \tfrac{1}{2}[\xi, \xi],
\end{align}
with shifted composite ghost defined by
\begin{align}
\label{shifted_dressed_ghost}
 \h{v}{\,}':=\h v+ i_\xi \h A.
\end{align}
Let us denote ${\h{\text{BRST}}}^\xi$ this algebra. Due to \eqref{shifted_dressed_ghost}  ${\h{\text{BRST}}}^\xi$ also assumes the second presentation, see~\eqref{mixed_BRST2},
\begin{align}
\label{shifted_dressed_BRST2}
\s \h A = -D\h v+ L_\xi \h A&, \quad \s \h F=[\h F, \h v\,] + L_\xi \h F, \quad \s \psi =-\rho_*(\h v\,) \h\psi + L_\xi \psi, \notag \\[2mm]
\s \h v&=-\tfrac{1}{2}[\h v, \h v\, ] +L_\xi \h v, \quad \s\xi= \tfrac{1}{2}[\xi, \xi].
\end{align}
The above form clearly shows the decoupling between the residual gauge symmetry ($\h v\,$) and the diffeomorphisms ($\xi$). 
\bigskip

The question now is to see whether the operations of shifting and dressing do commute, that is, whether $\h{\text{BRST}^\xi}$ \eqref{dressed_shifted_BRST} is the same as ${\h{\text{BRST}}}^\xi$ \eqref{shifted_dressed_BRST}. This is clearly the case if $\h{v'}={\h v}{\,}'$ and the latter is true if and only if the condition
\begin{align}
\label{A}
\s u= (s+L_\xi )\, u 
\end{align}
is satisfied. Indeed, 
\begin{align*}
\h{v'}&=u\- v'u+ u\-\s u  \stackrel{\eqref{A}}{=}   u\-(v+i_\xi A) u + u\-(s+L_\xi)u,\\
	&= \underbrace{u\-vu+u\-su}_\text{$=: \h v$} \ +  \underbrace{\ u\-i_\xi A u + u\-i_\xi du}_\text{$=i_\xi (u\- A u + u\-du)=i_\xi \h A$}
\end{align*}
since $i_\xi u=0$. Hence, one has proven that $\h{v'} = \h v +i_\xi \h A \rdefeq {\h v}{\,}'$. 
Conversely, $\h{v'}={\h v}{\,}'$ infers \eqref{A} as it can be easily shown. \footnote{ Notice that the second presentation of
  $\h{\text{BRST}^\xi}$ \eqref{dressed_shifted_BRST} is exactly
  \eqref{shifted_dressed_BRST2} when equation~\eqref{A} holds: the ghost mentioned in footnote~\ref{Second_Presentation} turns out to be $(u^{-1}vu+u^{-1}\s u) - u^{-1}L_\xi u=u^{-1}vu+u^{-1}su=:\h v$.} Symbolically, one may write [shifting, dressing]$=0$. 

On the other hand, if
\begin{align}
\label{noA}
\s u\neq  (s+L_\xi )\, u
\end{align}
then [shifting, dressing]$\neq0$. This can be summarized in the diagram: 
\vspace{-1.8cm}
\begin{center}
\includegraphics[scale=0.4]{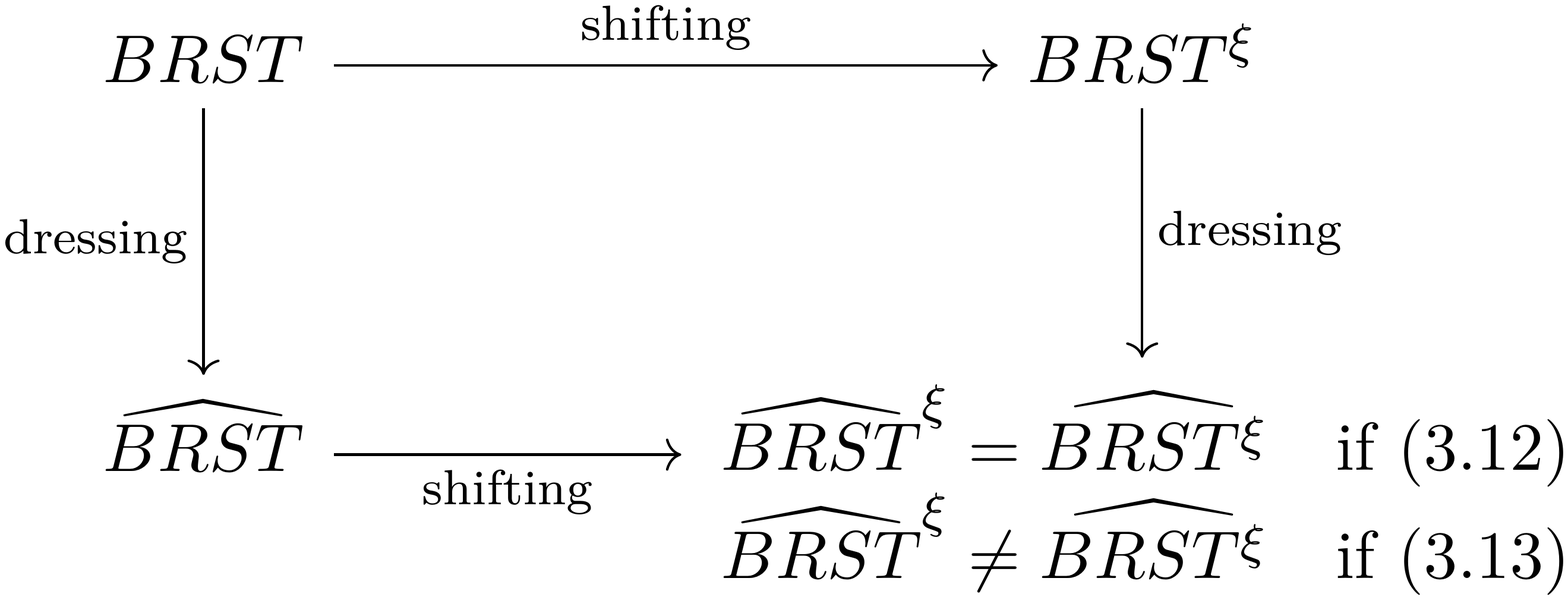}
\end{center}
\vspace{-2.3cm}
A criterion to decide in advance whether \eqref{A} or \eqref{noA}
holds is the following. By definition, the gauge transformation of a
dressing field $u$ is known and given by the $s$-operation. If
$u$ has no (free) spacetime index it is a $0$-form, so that its
transformation under $\Diff(\M)$ is actually given by the Lie derivative on
differential forms, $L_\xi= i_\xi d - d i_\xi$. Therefore, \eqref{A} holds in that case indeed. On the
contrary, if $u$ carries (free) spacetime indices, its transformation
under $\Diff(\M)$ is given by the Lie derivative $\L_\xi$ of tensors
\footnote{The reader is referred to~\cite[Chap.12]{Bertlmann}.}
or pseudo-tensors (as the case may be), and accordingly, \eqref{noA}
holds.
 
The two examples respectively treated in the next two
sections show that the differential algebra
which implements the correct infinitesimal gauge + diffeomorphism
symmetries will be $\h{\text{BRST}^\xi}$~\eqref{dressed_shifted_BRST}, that is the
one obtained by shifting first and then dressing. Notice that it provides the
most general form of the ghost~\eqref{dressed_shifted_ghost} which
takes into account the possible tensorial character of the dressing
field $u$ through the inhomogeneous term $u\- \sigma u$.
\smallskip

The two following examples concern the gauge formulation of pure gravitational theories (GR and conformal gravity), see {\em e.g.}~\cite{Blagojevic:2013xpa}. The natural language will be that of Cartan geometry~\cite{Sharpe, Kobayashi}, and the gravitational gauge potential is given by a (local) Cartan connection.

\section{Example: the geometry of General Relativity}  
\label{Example: the geometry of General Relativity}  

\subsection{Mixed Lorentz + $\Diff(\M)$ symmetry}  
\label{Mixed Lorentz  Diff(M) symmetry}  

The geometry of GR, seen as a gauge theory, is a Cartan geometry $(\P, \varpi)$ where $\P(\M, H)$ is a principal bundle with $H=SO(1, m-1)$ the Lorentz group, and $\varpi \in \Omega^1(\U, \LieG)$ is a (local) Cartan connection on $\U\subset \M$ with values in $\LieG$ the Lie algebra of the Poincaré group $G=SO(1, m-1) \ltimes \doubleR^{(1, m-1)}$.  One has the matrix representation of a gravitational field on spacetime $\M$, 
\begin{align*}
\varpi=\begin{pmatrix} A & \theta \\ 0 & 0 \end{pmatrix}=\begin{pmatrix} {A^a}_{b, \mu} & {e^a}_\mu \\ 0 & 0 \end{pmatrix}dx^\mu,
\end{align*}
with $A \in \Omega^1(\U, \LieH)$ the Lorentz connection (or spin connection) and $\theta\in \Omega^1(\U, \doubleR^{(1, m-1)})$ the soldering form (or vielbein $1$-form). The Greek indices are spacetime indices, while Latin indices are ``internal'' (gauge)-Minkowski indices. 
The curvature is 
\begin{align*}
\Omega=d\varpi+\tfrac{1}{2}[\varpi, \varpi]=d\varpi +\varpi \w \varpi \quad \rarrow \quad \begin{pmatrix} F & \Theta \\ 0 & 0 \end{pmatrix} := \begin{pmatrix} dA+A\w A & d\theta+A\w\theta \\ 0 & 0 \end{pmatrix},
\end{align*}
with $F$ the curvature $2$-form of $A$ and $\Theta$ the torsion $2$-form.
The Lorentz ghost  is
\begin{align*}
v=\begin{pmatrix} v_L & 0 \\ 0 & 0 \end{pmatrix} :\U \rarrow \LieH^*\otimes\LieH,
\end{align*}
and the associated BRST algebra reads
\begin{align*}
s\varpi&=-dv-[\varpi, v] \quad \rarrow\quad  \begin{pmatrix} sA & s\theta \\ 0 & 0 \end{pmatrix}=\begin{pmatrix} -Dv_L & -v_L\theta \\ 0 & 0 \end{pmatrix}:=\begin{pmatrix} -dv_L-[\varpi, v_L] & -v_L\theta \\ 0 & 0 \end{pmatrix} ,\\[2mm]
s\Omega&=[\Omega, v] \, \rarrow \,\begin{pmatrix} sF & s\Theta \\ 0 & 0 \end{pmatrix}= \begin{pmatrix} [F, v_L] & -v_L\Theta \\ 0 & 0 \end{pmatrix},\qquad
sv=-v^2 \, \rarrow \,  \begin{pmatrix} sv_L & 0 \\ 0 & 0 \end{pmatrix}= \begin{pmatrix} -v_L^2 & 0 \\ 0 & 0 \end{pmatrix}.
\end{align*}
This algebra handles the infinitesimal SO-gauge transformations of the variables of the theory. Defining the algebraic Cartan connection $\tilde \varpi:=\varpi +v$, one recovers the BRST algebra for GR from the horizontality condition $\tilde d \tilde \varpi +\tfrac{1}{2}[\tilde \varpi, \tilde \varpi]=\Omega$.

\medskip
Let us use the results of section~\ref{Adding diffeomorphisms} and write the shifted algebra $BRST^\xi$ for GR. The Lorentz ghost is shifted by the Cartan connection according to 
\begin{align*}
v'=v+i_\xi\varpi = \begin{pmatrix} v_L+i_\xi A & i_\xi\theta \\ 0 & 0 \end{pmatrix} =: \begin{pmatrix} v'_L & i_\xi\theta \\ 0 & 0 \end{pmatrix}
\end{align*}
and thus acquiring an effective ghost term $i_\xi\theta$ in the translation entry.
Hence, according to the general scheme given in section~\ref{The BRST gauge algebra},
we readily get
\[
\s \varpi = - Dv' + i_\xi\Omega = - Dv + L_\xi\varpi
\]
where $D = d + [A,\ ]$. In matrix notation it reads
\begin{align}
\label{sigma_connection}
 \begin{pmatrix} \s A & \s\theta \\ 0 & 0 \end{pmatrix}&=  \begin{pmatrix} -dv'_L & -d(i_\xi \theta) \\ 0 & 0 \end{pmatrix}-\left[ \begin{pmatrix} A & \theta \\ 0 & 0 \end{pmatrix},  \begin{pmatrix} v'_L & i_\xi\theta \\ 0 & 0 \end{pmatrix} \right] +  \begin{pmatrix} i_\xi F & i_\xi\Theta \\ 0 & 0 \end{pmatrix} \notag \\
 &=  \begin{pmatrix} -D v'_L + i_\xi F & -D(i_\xi\theta) - v'_L \theta + i_\xi\Theta\\ 0 & 0 \end{pmatrix}\\
 &= \begin{pmatrix} -Dv_L +L_\xi\varpi & -v_L\theta +L_\xi \theta \\ 0 & 0 \end{pmatrix}. \notag
\end{align}
Similarly, we have for the curvature
\begin{align*}
\s\Omega = [\Omega, v'] - D(i_\xi\Omega) = [\Omega, v] + L_\xi\Omega, \mbox{ \em i.e. }
\begin{pmatrix} \s F & \s\Theta \\ 0 & 0 \end{pmatrix}&=\begin{pmatrix} [F, v_L]+L_\xi F  & -v_L\Theta + L_\xi\Theta \\ 0 & 0 \end{pmatrix}
\end{align*}
and for the ghost field
\begin{align}
\label{sigma_Lorentz}
\s v' = -{v'}^2 + \tfrac{1}{2}i_\xi i_\xi \Omega,\quad \mbox{\em i.e.} \quad 
\begin{pmatrix} \s v'_L & \s i_\xi \theta \\ 0 & 0 \end{pmatrix} = \begin{pmatrix} -{v'_L}^2 + \tfrac{1}{2}i_\xi i_\xi F & -v'_L i_\xi \theta + \tfrac{1}{2}i_\xi i_\xi \Theta \\ 0 & 0 \end{pmatrix}\ .
\end{align}
One can also check that one recovers the presentation
\begin{align*}
\s v = -v^2 + L_\xi v, \quad \mbox{\em i.e.} \quad
\begin{pmatrix} \s v_L &0 \\ 0 & 0 \end{pmatrix} = \begin{pmatrix} -v_L^2+L_\xi v_L & 0 \\ 0 & 0 \end{pmatrix}.
\end{align*}
Formulas \eqref{sigma_connection} and \eqref{sigma_Lorentz} respectively reproduce equations (8a-b) and (8-c) of ``parallel transport'' given in~\cite{Langouche-Schucker-Stora} (once the background connection has been reabsorbed in the redefinition of the generators). See also~\cite{Stora:2005tp}.

\bigskip
The algebra $ \text{BRST}^\xi$ handles the full mixed symmetry Lorentz$+\Diff(\M)$ of GR. Now, thanks to the dressing field method, it is possible to reduce it so as to obtain a strict $\Diff(\M)$ algebra, in other words, to get the diffeomorphism symmetry only. 
 
\subsection{Residual $\Diff(\M)$ symmetry}  
\label{Residual Diff symmetry}  

As is detailed in \cite{GaugeInv, FLM15, JF_PhD}, the dressing field in GR is nothing but the vielbein,
$u:= \begin{pmatrix} e & 0 \\ 0 &  1 \end{pmatrix}$, $ e={e^a}_\mu$. 
The composite fields are, for the connection
\begin{align*}
\h\varpi = u^{-1}\varpi u + u^{-1}du = \begin{pmatrix} e^{-1} A e + e^{-1}de & e^{-1}\theta \\ 0 & 0 \end{pmatrix}
=: \begin{pmatrix} \Gamma & dx \\ 0 & 0 \end{pmatrix} = \begin{pmatrix} {\Gamma^\rho}_{\nu, \mu} & \delta^\rho_\mu \\ 0 & 0 \end{pmatrix}dx^\mu
\end{align*}
where the $\Gamma$'s are the Christoffel symbols of a metric connection for the metric $g=e^T \eta e$, while for the curvature
\begin{align*}
\h\Omega &= u^{-1}\Omega u = \begin{pmatrix} e^{-1}Fe & e^{-1}\Theta \\ 0& 0 \end{pmatrix}
=: \begin{pmatrix} R & T \\ 0 & 0 \end{pmatrix} = \tfrac{1}{2}\begin{pmatrix} {R^\rho}_{\nu, \mu\sigma} & {T^\rho}_{\nu, \mu\sigma}\\ 0 & 0 \end{pmatrix}dx^\mu\w dx^\sigma
\end{align*}
where $R$ and $T$ are the corresponding Riemann and torsion tensors, respectively.  
\medskip

Let us use the results of section \ref{Shifting and dressing} and write the algebra $\h{ \text{BRST}^\xi}$ for GR. The composite shifted ghost
is $\h{v'}=u\-v'u+u\-\s u$, which requires to know $\s u$ explicitly, that is $\s e$. 
The latter can be obtained from $\s\theta$, using the natural assumption that $\s x=0$, where $x$ is considered as a background system of local coordinates pertaining to the differentiable structure of the spacetime $\M$. One has
\begin{align}
\label{sigma_e}
\s\theta& = s\theta +L_\xi \theta,\notag \\
\s (e\cdot dx) &= s(e\!\cdot\! dx) + (i_\xi d - di_\xi) (e\!\cdot\! dx) = (se) \!\cdot\! dx + i_\xi(de \w dx) -d(e\!\cdot\! \xi),\notag\\
\s e\!\cdot\! dx &= (se +i_\xi de + e\!\cdot\!\d\xi ) \!\cdot\! dx.
\end{align}
Here ``$\cdot$'' is a shorthand for Greek index summation, e.g. $\theta =e\!\cdot\! dx:={e^a}_\mu dx^\mu$. One has then, 
\begin{align*}
\s u =   \setlength{\arraycolsep}{5pt}\begin{pmatrix} \s e & 0 \\ 0 &  0 \end{pmatrix}= \begin{pmatrix} se +i_\xi de + e\cdot\d\xi  & 0 \\ 0 &  0 \end{pmatrix}
\end{align*}
According to~\cite{Bertlmann} let us define $v_\xi =  \begin{pmatrix} \d \xi & 0 \\ 0 &  0 \end{pmatrix}$, and since $i_\xi u=0$, we finally obtain
\begin{align}
\label{sig_u_RG_noA}
\s u = su + L_\xi u + uv_\xi\ .
\end{align}
We are thus in a case where \eqref{noA} holds, so $\h{\text{BRST}^\xi}\neq \h{\text{BRST}}^\xi$ (the diagram does not commute). 
Remark that  $L_\xi u + uv_\xi:=\L_\xi u$  is the Lie derivative of the tensor $u\sim{e^a}_\mu$ (that is obvious since $L_\xi \theta=(\L_\xi {e^a}_\mu )dx^\mu $ ). This is a situation discussed at the very end of section \ref{Shifting and dressing}: the dressing $u$ has a free spacetime index and is a tensor, so its variation under $\Diff(\M)$ is indeed given by $\L_\xi$.  
The composite shifted Lorentz ghost is  thus
\begin{align*}
\h{v'}&=u\-v'u +u\-\s u = u\-(v+i_\xi\varpi)u + u\-(su +L_\xi u +uv_\xi) \\
	&=( u\-vu+u\-s u) \ + \ (u\-i_\xi\varpi u + u\-i_\xi du) \ + \ v_\xi \\
\h{v'}&= \h v + i_\xi \h\varpi + v_\xi.
\end{align*}
Of course, as expected $\h{v'}\neq \h v+i_\xi \h\varpi \rdefeq {\h v}{\,}'$. Moreover, in the situation at hand, we have from the initial Lorentz BRST algebra, $se=-v_Le$. Therefore, the composite Lorentz ghost vanishes,
\begin{align*}
\h v  =  u\-vu+u\-s u=\begin{pmatrix}  e\- & 0 \\ 0 &  1 \end{pmatrix}\begin{pmatrix} v_L & 0 \\ 0 &  0 \end{pmatrix}\begin{pmatrix}  e & 0 \\ 0 &  1 \end{pmatrix} + \begin{pmatrix}  e\- & 0 \\ 0 &  1 \end{pmatrix}\begin{pmatrix} se & 0 \\ 0 &  0 \end{pmatrix}=0.
\end{align*}
This means that the algebra $\h{\text{BRST}}$ is trivial in this case:
\begin{align*}
s\h\varpi= \begin{pmatrix} s\Gamma & sdx \\ 0 & 0 \end{pmatrix}=0, \qquad \text{and} \qquad s\h \Omega= \begin{pmatrix}s R & sT \\ 0 & 0 \end{pmatrix}=0\qquad \text{(and obviously $s\h v=0$)} .
\end{align*}
This expresses the Lorentz invariance of the composite fields. This is an instance of complete gauge neutralization as described in~\cite{ FLM15}. 

At last, the composite shifted Lorentz ghost is thus simply
\begin{align}
\label{hv'_RG}
\h{v'}&= i_\xi \h\varpi + v_\xi = \begin{pmatrix}  i_\xi\Gamma & i_\xi dx \\ 0 &  0 \end{pmatrix} + \begin{pmatrix}  \d\xi & 0 \\ 0 &  0 \end{pmatrix}
=\begin{pmatrix}  {\Gamma^\rho}_{\nu, \mu}\xi^\mu + \d_\nu \xi^\rho& \xi^\rho \\ 0 &  0 \end{pmatrix}
= \begin{pmatrix}  \nabla_\nu \xi^\rho& \xi^\rho \\ 0 &  0 \end{pmatrix}.
\end{align}
It is worth noticing that it depends \emph{covariantly} on the diffeomorphism ghost only. Straightforward matrix calculations now easily provide the algebra $\h{\text{BRST}^\xi}$. First,
\begin{align}
\label{sig_hvarpi_RG}
\s\h\varpi & = -D\h{v'} + i_\xi \h\Omega \notag\\
	      &= -d(i_\xi\h\varpi +v_\xi) -\cancel{[\h\varpi, i_\xi\h\varpi]} - [\h\varpi, v_\xi] + i_\xi d\h\varpi + \cancel{i_\xi\tfrac{1}{2}[\h\varpi, \h\varpi]} = (i_\xi d -di_\xi)\h\varpi -[\h\varpi, v_\xi] - dv_\xi \notag\\[2mm]
	     &= L_\xi \h\varpi -[\h\varpi, v_\xi] - dv_\xi
	     =:\L_\xi \h\varpi - dv_\xi\ .
\end{align}
In matrix form the last line reads
\begin{align*}
\s\h\varpi= \begin{pmatrix} \s\Gamma & \s dx \\ 0 &  0 \end{pmatrix} &=\setlength{\arraycolsep}{7pt}\begin{pmatrix}  L_\xi \Gamma -[\Gamma, \d\xi] -d\d\xi & L_\xi dx -\d\xi\!\cdot\! dx\\[2mm] 0 &  0 \end{pmatrix}\\[2mm]
	      &= \setlength{\arraycolsep}{6pt}\begin{pmatrix} \xi^\alpha \d_\alpha {\Gamma^\rho}_{\mu\nu} + {\Gamma^\rho}_{\alpha \nu} \d_\mu\xi^\alpha +  {\Gamma^\rho}_{\mu\alpha} \d_\nu\xi^\alpha - \d_\alpha\xi^\rho {\Gamma^\alpha}_{\mu\nu} + \d_\mu(\d_\nu\xi^\rho) & 0  \\[2mm] 0   & 0 \end{pmatrix}dx^\mu \\[2mm]
	      &=: \setlength{\arraycolsep}{7pt} \begin{pmatrix} \L_\xi {\Gamma^\rho}_{\mu\nu} + \d_\mu(\d_\nu\xi^\rho)  & 0 \\[2mm] 0  & 0 \end{pmatrix}dx^\mu 
\end{align*}
which gives the Lie derivative of the Christoffel symbols. Also
\begin{align}
\label{sig_hOmega_RG}
\s\h\Omega&=[\h\Omega, \h{v'}] - \h D(i_\xi\h\Omega) \notag\\
		 &=[\h\Omega, i_\xi\h\varpi]+ [\h\Omega, v_\xi] - di_\xi\h\Omega -[\h\varpi,  i_\xi\h\Omega]=i_\xi d\Omega - di_\xi\h\Omega + [\h\Omega, v_\xi] \notag\\
    		 &= L_\xi\h\Omega + [\h\Omega, v_\xi]=:\L_\xi \h\Omega
\end{align}
where in the course of the computation the Bianchi identity $i_\xi d\h\Omega= -i_\xi[\h\varpi, \h\Omega]$ has been used in the third equality.
In matrix notation one thus gets
\begin{align*}
\s \h\Omega= \begin{pmatrix}  \s R & \s T \\ 0 &  0 \end{pmatrix} = \begin{pmatrix}  L_\xi R + [R, \d\xi] & L_\xi T -\d\xi\!\cdot\! T \\ 0 &  0 \end{pmatrix} =  \tfrac{1}{2}\begin{pmatrix}   \L_\xi {R^\rho}_{\nu, \mu\sigma}& \L_\xi {T^\rho}_{ \mu\sigma} \\ 0 &  0 \end{pmatrix}dx^\mu \w dx^\sigma
\end{align*}
from which one reads the well-known Lie derivatives of the Riemann and torsion tensors respectively
\begin{align}
\label{R_and_T}
\L_\xi {R^\rho}_{\nu, \mu\sigma} &= \xi^\alpha \d_\alpha {R^\rho}_{\nu, \mu\sigma} + {R^\rho}_{\nu, \alpha\sigma} \d_\mu\xi^\alpha + {R^\rho}_{\nu, \mu\alpha} \d_\sigma\xi^\alpha + {R^\rho}_{\alpha, \mu\sigma} \d_\nu\xi^\alpha - \d_\alpha \xi^\rho {R^\alpha}_{\nu, \mu\sigma}  \notag\\[-3mm]
& \\[-4mm]
\L_\xi {T^\rho}_{\mu\sigma} &= \xi^\alpha \d_\alpha {T^\rho}_{ \mu\sigma} + {T^\rho}_{ \alpha\sigma} \d_\mu\xi^\alpha + {T^\rho}_{ \mu\alpha} \d_\sigma\xi^\alpha - \d_\alpha \xi^\rho {T^\alpha}_{\mu\sigma} \ . \notag
\end{align}
At this stage, since we know that $\s^2=0$  on $\h\varpi$ (and $\h\Omega$) requires \eqref{sig_xi0},
\footnote{Indeed this does not depend on form of the shifted ghost. 
}
the action of $\s$ on the relevant variables $\h\varpi$, $\h\Omega$ and $\xi$ ($\h v$ being vanishing) is  known. But, for the sake of completeness, we nevertheless write the last relation
\begin{align*}
\s\h{v'} &= -\tfrac{1}{2}[\h{v'}, \h{v'}] +\tfrac{1}{2}i_\xi i_\xi \h\Omega \\
	   &= L_\xi i_\xi\h\varpi - [i_\xi\h\varpi, v_\xi] + L_\xi v_\xi -\tfrac{1}{2}[v_\xi, v_\xi] - i_\xi dv_\xi - i_{\tfrac{1}{2}[\xi, \xi]}\h\varpi \\
	&= \L_\xi\h{v'} - i_\xi dv_\xi - i_{\tfrac{1}{2}[\xi,\xi]} \h\varpi \ . 
\end{align*}
This is redundant with \eqref{sig_hvarpi_RG} and provides only an indirect checking on the variation $\s\xi$ to secure that $\s^2\h{v'}=0$.
\medskip

The algebra $\h{ \text{BRST}^\xi}$ thus gives the correct transformations of
the Christoffel symbols, the Riemann and the torsion tensors
under infinitesimal diffeomorphisms. Since in this case the
Lorentz-gauge symmetry is neutralized, the shifted algebra $\h{ \text{BRST}^\xi}$ handles the spacetime symmetry only as a residual symmetry.

\section{Example: the second-order conformal structure}  
\label{Example: the second-order conformal structure}  

The joint treatment of Weyl symmetry and diffeomorphisms within the BRST framework has been addressed (algebraically) by several authors, see {\em e.g.}~\cite{Bonora:1984pz,Bonora:1984ic,Bonora-Pasti-Bregola:1986,Moritsch:1994hv} or later on  \cite{Boulanger1,Boulanger2,Boulanger:2007st}. 
We shall use here the geometrical view that the Weyl symmetry is involved in the second order conformal structure which is well described in the framework of a Cartan geometry. We shall
apply the scheme depicted in section~\ref{The dressing field method and diffeomorphisms} to the corresponding BRST algebra.

We refer the reader to \cite{Sharpe} and to \cite{Kobayashi, Ogiue} for mathematical details. Here, we just sketch the necessary material to follow our scheme, but we also heavily rely on results detailed in~\cite{FLM15}.
\medskip

The whole structure is modeled on the Klein pair of Lie groups $(G, H)$  where $G$  is the Möbius group and $H$ is the subgroup such that $G/H \simeq (S^{m-1}\times S^1) / \mathbb{Z}^2$ (the compactified Minkoswki space considered as homogeneous space, see {\em e.g.} \cite{Go:1971zz}) and has the following factorized matrix presentation
\begin{align}
\label{HKK}
 H = K_0\, K_1=\left\{ \begin{pmatrix} z &  0 & 0  \\  0  & S & 0 \\ 0 & 0 & z^{-1}  \end{pmatrix}\!  \begin{pmatrix} 1 & r & \frac{1}{2}rr^t \\ 0 & \1 & r^t \\  0 & 0 & 1\end{pmatrix}  \bigg|\, z\in \mathsf{W}=\doubleR^*_+,\ S\in SO(1, m-1), 
\ r\in \doubleR^{m*} \right\}.
\end{align} 
Here ${}^t$ stands for the $\eta$-transposition, namely for the row vector $r$ one has $r^t = (r \eta^{-1})^T$ (the operation ${}^T\,$ being the usual matrix transposition), and $\doubleR^{m*}$ is the dual of $\doubleR^m$. 
We refer to $\mathsf{W}$ as the Weyl group of rescaling. Obviously $K_0\simeq CO(1, m-1)$, and $K_1$ is the abelian group of inversions (or special conformal transformations). 

Infinitesimally, we have the Klein pair  $(\LieG, \LieH)$ of graded Lie algebras~\cite{Kobayashi}. They decompose respectively as, $\LieG=\LieG_{-1}\oplus\LieG_0\oplus\LieG_1 \simeq \doubleR^m\oplus\co(1, m-1)\oplus\doubleR^{m*}$, a splitting which gives the different sectors of the conformal rigid symmetry: translations + (Weyl $\times$ Lorentz) + inversions, and $\LieH=\LieG_0\oplus\LieG_1 \simeq \co(1,m-1)\oplus\doubleR^{m*}$. In matrix notation we have,
\begin{align*}
\mathfrak{g} = \left\{ 
\begin{pmatrix} \epsilon &  \iota & 0  \\  \tau  & v & \iota^t \\ 0 & \tau^t & -\epsilon  \end{pmatrix} \bigg|\ (v-\epsilon\1)\in \mathfrak{co},\ \tau\in\mathbb{R}^m,\ \iota\in\mathbb{R}^{m*}  
\right\} 
\supset
\LieH = \left\{ \begin{pmatrix} \epsilon &  \iota & 0  \\  0  & v & \iota^t \\ 0 & 0 & -\epsilon  \end{pmatrix} \right\},
\end{align*} 
with the $\eta$-transposition $\tau^t = (\eta\tau)^T$ of the  column vector $\tau$.
The graded structure of the Lie algebras, $[\LieG_i, \LieG_j] \subseteq \LieG_{i+j}$, $i,j=0,\pm 1$ with the abelian Lie subalgebras $[\LieG_{-1}, \LieG_{-1}] = 0 = [\LieG_1, \LieG_1]$, is automatically handled by the matrix commutator.
\medskip

One can thus ``localized'' the conformal group by considering the second-order conformal structure. The latter is a Cartan geometry $(\P, \varpi)$ where $\P=\P(\M, H)$ is a principal bundle over $\M$ with structure group $H=K_0 K_1$, and $\varpi\in \Omega^1(\U, \LieG)$ is a  (local) Cartan connection. The curvature is given by $\Omega=d\varpi +\tfrac{1}{2}[\varpi, \varpi]=d\varpi +\varpi^2$. Both have a matrix representation 
\begin{align*}
\varpi=\begin{pmatrix} a & \alpha & 0 \\ \theta & A & \alpha^t \\ 0 & \theta^t & -a   \end{pmatrix} \qquad \text{and} \qquad \Omega=\begin{pmatrix} f & \Pi & 0  \\ \Theta & F & \Pi^t \\ 0 & \Theta^t & -f \end{pmatrix}.
\end{align*}
One can single out the so-called normal conformal Cartan connection
(which is unique) by imposing the constrains $\Theta=0$ (torsion free)
and ${F^a}_{bad}=0$. Together with the $\LieG_{-1}$-sector of the
Bianchi identity, $d\Omega+[\varpi, \Omega]=0$, these imply $f=0$ (trace
free), so that the curvature of the normal Cartan connection reduces
to 
\[
\Omega = \begin{pmatrix}
0 & \Pi & 0  \\ 0 & F & \Pi^t \\[0.5mm] 0
  & 0 & 0 \end{pmatrix} \qquad \mbox{(normal case)}.
\]
In the normal geometry, $\alpha$ is the Schouten $1$-form, $\Pi$ and $F$ are the Cotton and Weyl $2$-forms respectively. 

The ghost field, $v:\U \rarrow \LieH^*\otimes\LieH$, associated with the $\H$-gauge symmetry is given by
\begin{align*}
v=v_{\mathsf{W}}+v_L+v_i=\begin{pmatrix} \epsilon &  0 & 0 \\ 0 & 0 &0 \\ 0 & 0& -\epsilon\end{pmatrix} + \begin{pmatrix} 0 &  0 & 0 \\ 0 & v_L &0 \\ 0 & 0& 0 \end{pmatrix}+\begin{pmatrix} 0 &  \iota & 0 \\ 0 & 0 & \iota^t \\ 0 & 0& 0 \end{pmatrix}.
\end{align*}
With this matrix representation, the associated BRST algebra reads as usual
\begin{align}
\label{BRST_conf}
s\varpi=-Dv:=-dv - [\varpi, v], \qquad s\Omega=[\Omega, v], \quad \text{and} \quad sv=-\tfrac{1}{2}[v, v]=-v^2.
\end{align}

Defining the corresponding algebraic conformal Cartan connection $\tilde \varpi:=\varpi+v$, the algebra~\eqref{BRST_conf} is  
recovered from the Russian formula, $\tilde d \tilde \varpi + \tfrac{1}{2}[\tilde\varpi, \tilde\varpi]=\Omega$. 
\medskip 

The principal bundle $\P(\M, H)$ is a second order $G$-structure, a reduction of the second order frame bundle $L^2\M$; it is thus a ``2-stage bundle''. The bundle $\P(\M,H)$ over $\M$ can also be seen as a principal bundle $\P_1:=\P(\P_0,K_1)$ with structure group $K_1$ over the principal bundle $\P_0:=\P(\M,K_0)$.

Accordingly, in \cite{FLM15} we showed how (locally) the structure group $H$ could be reduced in two steps: first from $H$ to $K_0$ by neutralizing $K_1$ with a first dressing field $u_1$, then from $K_0$ to $\mathsf{W}$ thanks to a second dressing field $u_0$. We displayed the corresponding sequence of reduced BRST algebras \footnote{Due to the successive dressings the $\widehat{\ }$ is dropped out to the benefit of a lower index.}
\begin{align*}
\text{BRST}: (\varpi,\ \Omega,\ s,\ v ) \xrarrow{u_1} \text{BRST}_1: (\varpi_1,\ \Omega_1,\ s,\ v_1) \xrarrow{u_0} \text{BRST}_0: (\varpi_0,\ \Omega_0,\ s,\ v_0).
\end{align*}
We also showed that it is possible to define $u:=u_1u_0$ and to proceed in a single step, 
\begin{align*}
\text{BRST}: (\varpi,\ \Omega,\ s,\ v ) \xrarrow{u} \text{BRST}_0: (\varpi_0,\ \Omega_0,\ s,\ v_0).
\end{align*}

\subsection{Shifting and first dressing field}   
\label{Shifting and the first dressing field}   

The shift of the differential algebra~\eqref{BRST_conf} in order to include infinitesimal diffeomorphisms is performed as in the general case:
\begin{align*}
\text{BRST}: \big(\varpi,\ \Omega,\ s,\ v \big) \quad \xrarrow{\xi} \quad \text{BRST}^\xi :  \big(\varpi,\ \Omega,\ \s,\ 
v'= v+i_\xi\varpi \big),
\end{align*}
and due to the decomposition of the ghost, $v'=v+i_\xi\varpi$, we know that $\s =s+L_\xi$ on $(\varpi, \Omega, v)$.
\medskip

It is interesting to see what happens under the first dressing operation by $u_1$. The latter gives the sequence,
\begin{align*}
{BRST^\xi}:  \big(\varpi,\ \Omega,\ \s,\ v' \big) \quad \xrarrow{u_1} \quad 
{(BRST^\xi)_1}:  \big(\varpi_1,\ \Omega_1\ , \s,\ (v')_1\!:=u_1\- v' u_1 + u_1\- \s u_1 \big).
\end{align*}
The crux is of course to determine $\s u_1$. In \cite{FLM15} we defined the dressing field $u_1:\U \rarrow K_1$ by
\begin{align*}
u_1 :=\begin{pmatrix} 1 & q & \tfrac{1}{2}qq^t \\ 0 & \1 & q^t \\ 0 & 0 & 1  \end{pmatrix}  \qquad \text{with}\quad  q:= a\!\cdot\! e\- \in \doubleR^{(1, m-1)*},  
\end{align*}
where $\bs a = a\!\cdot\! dx$ and $\theta = e\!\cdot\! dx$ (with indices, $q_a:= a_\mu {(e\-)^\mu}_a$). Hence,
\begin{align*}
\s u_1\sim \s q=\s(a\cdot e\-)=(\s a) \cdot e\- - a \cdot e\-(\s e)e\-,
\end{align*}
and we need to know $\s a$ and $\s e$, that is $\s \varpi$. Since $\s
\varpi=s \varpi + L_\xi \varpi$, one has
\begin{align*}
\s \bs a = (s + L_\xi )\bs a, \qquad \text{and} \qquad \s \theta = (s+L_\xi)\theta.
\end{align*}
The first equation reads, 
\begin{align*}
\s (a\!\cdot\! dx) &= s(a\!\cdot\! dx) + (i_\xi d - di_\xi) (a\!\cdot\! dx) = (sa) \!\cdot\! dx + i_\xi(da \w dx) - d(a\!\cdot\! \xi) \\
		    &=(sa) \!\cdot\! dx + i_\xi da \!\cdot\! dx +\cancel{da\!\cdot\!\xi} - \cancel{da\!\cdot\!\xi} -a\!\cdot\! d\xi.
\end{align*}
This gives, $ \s a = sa + i_\xi da + a\!\cdot\! \d\xi$. 
The second equation is already known from \eqref{sigma_e}.
Finally, 
\begin{align*}
\s q &= (\s a)\!\cdot\! e\- -a\!\cdot\! e\-(\s e)e\-,\\
       &=(sa + i_\xi da + a\!\cdot\! \d\xi)\!\cdot\! e\-  
-a\!\cdot\! e\-(se + i_\xi de + e\!\cdot\! \d\xi)e\-,\\
       &= (sa)\!\cdot\! e\- +i_\xi da \!\cdot\! e\-  + \cancel{(a\!\cdot\! \d\xi)\!\cdot\! e\-} + a\!\cdot\! se\- + a\cdot i_\xi de\- \cancel{-(a\!\cdot\! \d \xi)\!\cdot\! e\-},\\
       &=s(a\!\cdot\! e\-) + i_\xi d(a\!\cdot\! e\-),\\
       &= sq + i_\xi dq .
\end{align*}
Noticing that $i_\xi q=0$, we end up with the result, 
\begin{align}
\label{sig_u_1}
\s q = (s+L_\xi)q \quad \mbox{so that} \quad \s u_1 = (s + L_\xi) u_1
\end{align}
This shows that the first dressing of the conformal structure satisfies \eqref{A}. Therefore, from our general discussion of section \ref{Shifting and dressing}  we can conclude that,
\begin{align}
\label{v'_1}
(v')_1 = u_1\- v' u_1 + u_1\- \s u_1 = v_1 + i_\xi \varpi_1 \rdefeq (v_1)'.
\end{align}
This means that $(\text{BRST}^\xi)_1 = (\text{BRST}_1)^\xi$, and we have the commutative diagram
\begin{center}
\begin{tikzcd}[column sep=huge, row sep=large]
\text{BRST} \arrow{r}{\xi} \arrow{d}[swap]{u_1}
& \text{BRST}^\xi \arrow{d}{u_1} \\
\text{BRST}_1 \arrow{r}{\xi}
& (\text{BRST}_1)^\xi = (\text{BRST}^\xi)_1   
\end{tikzcd}  
\end{center}
The ghost $(v')_1=(v_1)'$ encodes the residual symmetry $CO(1, m-1)+\Diff(\M)$ and the reduced shifted algebra $(\text{BRST}^\xi)_1=(\text{BRST}_1)^\xi$ handles the transformation of the composites fields $\varpi_1$ and $\Omega_1$ under these symmetries. In its second, decoupled, presentation it reads
\begin{align*}
\s \varpi_1 = -D v_1  +L_\xi \varpi_1 ,\qquad \s\Omega_1=[\Omega_1, v_1] + L_\xi\Omega_1 \qquad \text{and} \qquad \s v_1=-\tfrac{1}{2}[v_1, v_1] + L_\xi v_1.
\end{align*}
We refer to \cite{FLM15} for the detailed results concerning the algebra $\text{BRST}_1$ associated to the first composite ghost $v_1$. 

As already mentioned, it is possible to further reduce the gauge symmetry with a second dressing field $u_0$.

\subsection{Shifting and the second dressing field}
\label{Shifting and the second dressing field} 

We here face a situation which is analogous to the GR case, the second dressing field $u_0$ we now use is the vielbein, extracted from the Cartan connection $\varpi$. 
Again, from general results of section~\ref{Shifting and dressing}, we know that upon dressing $(\text{BRST}^\xi)_1$ with  $u_0$ we have the change of differential algebras
\begin{align*}
{(\text{BRST}^\xi)_1}\!:  \big(\varpi_1, \Omega_1, \s, (v')_1 \big)  \xrarrow{u_0}  {(\text{BRST}^\xi)_{1,0}}\!:  \big(\varpi_0, \Omega_0 , \s, (v')_{1,0}\!:=u_0\- (v')_1 u_0\!+\!u_0\- \s u_0 \big)
\end{align*}
whose outcoming ghost reads,
\begin{align}
\label{v'_1_0}
(v')_{1,0}&=u_0\-(v')_1 u_0 + u_0\-\s u_0 = u_0\-\big( u_1\- v' u_1 + u_1\-\s u_1\big) u_0 +  u_0\-\s u_0,\nonumber\\
	      &= (u_1u_0)\- v' (u_1u_0) + (u_1u_0)\-\s(u_1u_0),\nonumber\\[2mm]
(v')_{1,0}&=u\-v'u +u\- \s u, \qquad \text{by defining } u\defeq u_1u_0.
\end{align}
Equation \eqref{v'_1_0} shows that one can start from $\text{BRST}^\xi$ and use the dressing field $u=u_1u_0$ to obtain the algebra $(\text{BRST}^\xi)_{1,0}$ in a single step. This is possible because the two dressing fields satisfies the following compatibility conditions 
\begin{align*}
 u_1^S=S\-u_1S, \quad \text{and} \quad u_0^{\gamma_1}=u_0
\end{align*}
regarding their transformations under the gauge subgroups $\SO$ and $\K_1$. These, together with the defining transformations properties  $u_1^{\gamma_1}=\gamma_1\-u_1$ and $u_0^S=S\-u_0$, entail that $u$ is indeed a dressing under the product~$\SO\, \K_1 \subset \H$. 
\bigskip 

 In an alternative but equivalent way, upon dressing $(\text{BRST}_1)^\xi$ we have the change of differential algebras
 \begin{align*}
{(\text{BRST}_1)^\xi}\!:  \big(\varpi_1, \Omega_1, \s, (v_1)' \big) \xrarrow{u_0}  {[(\text{BRST}_1)^\xi ]_0}\!: \! \big(\varpi_0, \Omega_0 , \s, [(v_1)']_0\!:=\!u_0\- (v_1)' u_0\!+\! u_0\- \s u_0 \big)
\end{align*}
whose outcoming  ghost reads,
\begin{align}
\label{v_1'_0}
[(v_1)']_0 &= u_0\-\big( v_1 +i_\xi\varpi_1\big)u_0 + u_0\-\s u_0 \notag\\
	      &= u_0\-\big( u_1\-vu_1+u_1\-s u_1 + u_1\-i_\xi \varpi u_1 + u_1\-i_\xi du_1 \big)u_0 + u_0\-\s u_0
	      \notag\\[2mm]
[(v_1)']_0 &= u\-\big( v+ i_\xi \varpi \big)u + u\- \big(su_1+i_\xi du_1\big)u_0 +u_0\- \s u_0 \ .
\end{align}

Of course, both \eqref{v'_1_0} and \eqref{v_1'_0} are the same, and in any case the key question is  the value of $\s u_0$. But  the answer is already at hand since we know that $\s e = se + i_\xi de + e\!\cdot\! \d\xi$. With the definitions  
\begin{align*}
u_0:=\begin{pmatrix} 1 & 0 & 0 \\ 0 & e &0 \\ 0 & 0& 1 \end{pmatrix}
\text{ given in \cite{FLM15}, and }  v_\xi:=\begin{pmatrix} 0 & 0 &0 \\ 0 & \d \xi & 0 \\ 0& 0& 0 \end{pmatrix}
\end{align*}
we have  then
\begin{align}
\label{sig_u_0}
\s u_0 = su_0 + L_\xi u_0 + u_0 v_\xi
\end{align}
which is an instance of \eqref{noA}. This means that the final ghost has the decomposition
\begin{align*}
(v')_{1,0}=[(v_1)']_0
	     = u\-vu + u\-su + i_\xi (u\-\varpi u+u\-d u) + v_\xi \ .
\end{align*}
One can easily recognize the final Weyl ghost $v_0:=u\-vu + u\-su $, and the final composite field $\varpi_0:=u\-\varpi u+u\-d u$, obtained in a single step upon dressing $v$ and $\varpi$ with $u:=u_1u_0$.
By simply denoting ${v'}_0\defeq (v')_{1,0}=[(v_1)']_0$, we have
\begin{align}
\label{v'_0}
{v'}_0= v_0 + i_\xi \varpi_0 + v_\xi \ .
\end{align}
As expected ${v'}_0\neq v_0+i_\xi\varpi_0=(v_0)'$, and we can sum up the steps in the following diagram
\begin{center}
\begin{tikzcd}[column sep=huge, row sep=large]
\text{BRST} \arrow{r}{\xi} \arrow{d}[swap]{u_1}
& \text{BRST}^\xi \arrow{d}{u_1}
\arrow[bend left=77]{dd}{u = u_1 u_0} \\
\text{BRST}_1 \arrow{r}{\xi} \arrow{d}[swap]{u_0}
& (\text{BRST}_1)^\xi = (\text{BRST}^\xi)_1 \arrow{d}{u_0} \\ 
\text{BRST}_{1,0} \arrow[dashed]{r}{\xi}
& \mbox{$[(\text{BRST}_1)^\xi]_0$} = (\text{BRST}^\xi)_{1, 0}
\end{tikzcd}  
\end{center}
On the right hand side, the curved arrow illustrates the implication of \eqref{v'_1_0}. The bottom dashed arrow indicates that the diagram does not close there, as is clear from \eqref{v'_0}, consequence of \eqref{sig_u_0}.
\bigskip

As an illustration of the criterion discussed at the end of section \ref{Shifting and dressing}, let us remark that \eqref{A} holds for $u_1\sim q_a=a_\mu {(e\-)^\mu}_a$ since it has only a free internal index and no free spacetime index. It is a $0$-form  and is thus scalar under coordinate changes. Whereas \eqref{noA} holds for $u_0 \sim {e^a}_\mu$ since it carries both an internal index \textit{and} a free spacetime index. Thus it is tensorial (covector) under coordinate changes.
 \bigskip

Using the matrix form of the Weyl ghost $v_0$ and of the composite field $\varpi_0$ found in \cite{FLM15}, the final ghost \eqref{v'_0} reads explicitly
\begin{align*}
{v'}_0=\begin{pmatrix} \epsilon & \d\epsilon +P\!\cdot\!\xi & 0 \\[0.25em] \xi & \epsilon\delta + \nabla\xi & g\-\!\cdot\!\big(\d\epsilon + \xi P^T \big) \\[0.25em] 0 & \xi\!\cdot\! g & -\epsilon \end{pmatrix}=
\begin{pmatrix} \epsilon & \d_\nu\epsilon + P_{\nu\lambda}\xi^\lambda  & 0 \\[0.25em] \xi^\rho & \epsilon\delta^\rho_\nu\ +\ \d_\nu\xi^\rho +{\Gamma^\rho}_{\nu\lambda}\xi^\lambda   & g^{\rho\alpha}\big(\d_\alpha\epsilon + \xi^\lambda P_{\lambda\alpha} \big)\\[0.25em] 0 & \xi^\lambda g_{\lambda\nu} & -\epsilon \end{pmatrix}
\end{align*}
 where $g =e^T\eta e$ is the metric, $\Gamma$ is a linear connection and $P$ is a generalization of the Schouten tensor.
The final ghost encodes in a covariant way the residual mixed symmetry Weyl+$\Diff(\M)$ with generators $(\epsilon,\xi)$. 

It results that the composite shifted algebraic connection, $\varpi_0 + {v'}_0$,
gives a geometrical interpretation to the results obtained in \cite{Boulanger1}. Indeed, in this paper which aims at constructing a Weyl-covariant tensor calculus, the entries of $ \varpi_0 + {v'}_0$ are found as fields  (called \textit{generalized connections}) belonging to a space of variables identified through BRST-cohomological techniques. To some extent, one ought to say that the BRST cohomologies capture pieces of the geometry.
Now, it is easy to write the final algebra $(\text{BRST}^\xi)_{1, 0}$
\begin{align*}
\s\varpi_0=-D_0{v'}_0 + i_\xi \Omega_0, \quad \s\Omega_0= [\Omega_0, {v'}_0] \!-\! D_0(i_\xi\Omega_0),\quad \text{and}\quad  \s{v'}_0=-\tfrac{1}{2}[{v'}_0, {v'}_0] + \tfrac{1}{2}i_\xi i_\xi\Omega_0.
\end{align*}
On account of the decomposition \eqref{v'_0} of ${v'}_0$ we have first, 
\begin{align}
\label{sig_varpi_0}
\s \varpi_0 &= -dv_0 - di_\xi\varpi_0 - dv_\xi - [\varpi_0, v_0] - \cancel{ [\varpi_0, i_\xi\varpi_0]} - [\varpi_0, v_\xi] + i_\xi d\varpi_0+ \cancel{\tfrac{1}{2} i_\xi[\varpi_0, \varpi_0]} \notag\\
		&=  -dv_0 -  [\varpi_0, v_0] + L_\xi \varpi_0 - [\varpi_0, v_\xi] - dv_\xi \notag \\
\s \varpi_0 &= s_\mathsf{W}\varpi_0 + \L_\xi \varpi_0 - dv_\xi,
\end{align}
 where $s_\mathsf{W}$ is the Weyl BRST operator associated with the Weyl ghost $v_0$. Next we have, 
 \begin{align*}
  \s\Omega_0 &= [\Omega_0, v_0] + [\Omega_0, i_\xi\varpi_0] +
 [\Omega_0, v_\xi] - di_\xi\Omega_0 - [\varpi_0, i_\xi\Omega_0]
\\
 		   &= [\Omega_0, v_0] - i_\xi [\varpi_0, \Omega_0] + [\Omega_0, v_\xi] - di_\xi\Omega_0 = [\Omega_0, v_0] + L_\xi \Omega_0 + [\Omega_0, v_\xi]
 \end{align*}
 where in the third equality, by the Bianchi identity, $i_\xi d\Omega_0+i_\xi [\varpi_0, \Omega_0]=0$ has been used. One thus gets
\begin{equation}
\label{sig_Omega_0}
\s\Omega_0 =  s_\mathsf{W}\Omega_0 + \L_\xi \Omega_0 \ .
\end{equation}
Finally, with a little bit more of effort one finds
 \begin{align}
 \label{sigma_v'_0}
 \s {v'}_0&= (s_\mathsf{W} +\L_\xi){v'}_0 - i_\xi dv_\xi - i_{\tfrac{1}{2}[\xi, \xi]}\varpi_0 \ .
 \end{align}
 Most part of this equation is redundant with \eqref{sig_varpi_0} and
 the fact that  $\s \xi=\tfrac{1}{2}[\xi, \xi]$, implied by
 $\s^2=0$. But the Weyl subalgebra part, $s_\mathsf{W} v_0$, expresses
 the fact that the Weyl group of scale transformations is abelian: $s_\mathsf{W} \epsilon=0$. 
 Equations \eqref{sig_varpi_0} and \eqref{sig_Omega_0} express the transformation laws of the composite fields $\varpi_0$ and $\Omega_0$ under the mixed symmetry Weyl+$\Diff(\M)$. 

\subsection{The normal case}

In the instance of the normal conformal Cartan connection, the normality conditions on the curvature are preserved through the successive dressing operations~\cite{FLM15} and the final composite fields are
 \begin{align*}
 \varpi_0 = \begin{pmatrix} 0 & P & 0 \\[1mm] dx & \Gamma & g\-\!\cdot\!P^T \\[1mm] 0 & dx^T\!\cdot\!\! g & 0\end{pmatrix}= \begin{pmatrix} 0 & P_{\mu\nu} & 0 \\[1mm]  \delta_\mu^\rho &\ {\Gamma^\rho}_{\mu\nu}& \ g^{\rho\lambda}P_{\lambda\mu} \\[1mm] 0 & g_{\mu\nu} & 0 \end{pmatrix} dx^\mu
\end{align*}
where $g$ is the metric tensor, $\Gamma$ is the Levi-Civita connection, $P$ is the Schouten tensor (expressed in terms of the Ricci tensor and Ricci scalar)
\[
P_{\mu\nu} = \frac{-1}{(m-2)} \bigg( R_{\mu\nu} - \frac{R}{2(m-1)} g_{\mu\nu}\!\bigg)
\]
and for the curvature
 \begin{align*}
 \Omega_0 = \begin{pmatrix} 0 & C & 0 \\[1mm] 0 & W & g\-\!\cdot\!C^T \\[1mm] 0 & 0 & 0\end{pmatrix}=\tfrac{1}{2}\begin{pmatrix} 0 & C_{\nu, \mu\sigma} & 0 \\[1mm]  0 & \ {W^\rho}_{\nu, \mu\sigma} & g^{\rho\lambda}C_{\lambda, \mu\sigma} \\[1mm] 0 & 0 & 0 \end{pmatrix} dx^\mu\wedge dx^\sigma
\end{align*}
where $C$ and $W$ are the Cotton and Weyl tensors respectively. This is known as the \emph{Riemannian parametrization of the normal conformal Cartan connection}. 
 
The Weyl subalgebra $s_\mathsf{W}\varpi_0$ and $s_\mathsf{W}\Omega_0$ of  $(\text{BRST}^\xi)_{1, 0}$ then easily gives, through a simple matrix calculation, the transformations of the above mentioned various objects under infinitesimal Weyl rescaling. This was detailed in \cite{FLM15}. Instead, let us give the explicit matrix form of the $\Diff(\M)$ subalgebra in \eqref{sig_varpi_0} and \eqref{sig_Omega_0}. Again this is simple matrix calculations. 

First, let us compute the combination 
\begin{align}
 \label{expl_sig_varpi_0}
\MoveEqLeft{\L_\xi \varpi_0 - dv_\xi = L_\xi \varpi_0 - [\varpi_0, v_\xi] -d v_\xi}  \notag\\[3mm]
& = L_\xi \begin{pmatrix}
0 & P & 0 \\
dx & \Gamma & g\-\!\cdot\! P^T \\
0 & dx^T\!\cdot\!g & 0 
\end{pmatrix} 
-\left[
\begin{pmatrix} 0 & P & 0 \\ dx & \Gamma & g\-\!\cdot\! P^T \\ 0 & dx^T\!\cdot\!g & 0 \end{pmatrix}
\!,\!
\begin{pmatrix} 0 & 0 & 0 \\ 0 & \d\xi & 0 \\ 0 & 0 & 0 \end{pmatrix}
\right]
- \begin{pmatrix} 0 & 0 & 0 \\ 0 & d\d\xi & 0 \\ 0 & 0 & 0 \end{pmatrix} \notag\\[3mm]
& = \begin{pmatrix}
0 & L_\xi P - P\cdot \d\xi & 0 \\[2mm]
0 & L_\xi\Gamma - [\Gamma, \d\xi] - d\d\xi & L_\xi(g\-\!\cdot\!P^T) - \d\xi\cdot(g\-\!\cdot\!P^T) \\[2mm]
0 & L_\xi(dx^T\!\cdot\!g) - (dx^T\!\cdot\!g)\cdot \d\xi & 0
\end{pmatrix} \notag\\[3mm]
& = 
\begin{pmatrix}
0 & \xi^\alpha \d_\alpha P_{\mu\nu} + P_{\alpha\nu} \d_\mu \xi^\alpha + P_{\mu\alpha} \d_\nu \xi^\alpha & 0 \\[2mm]
0 & \xi^\alpha \d_\alpha {\Gamma^\rho}_{\mu\nu} + {\Gamma^\rho}_{\alpha \nu} \d_\mu\xi^\alpha +
{\Gamma^\rho}_{\mu\alpha} \d_\nu\xi^\alpha - \d_\alpha\xi^\rho {\Gamma^\alpha}_{\mu\nu} + \d_\mu(\d_\nu\xi^\rho) & * \\[2mm]
0 & \xi^\alpha \d_\alpha g_{\mu\nu} + g_{\alpha\nu} \d_\mu \xi^\alpha + g_{\mu\alpha} \d_\nu \xi^\alpha & 0
\end{pmatrix}\ dx^\mu \notag\\[3mm]
& =:
\begin{pmatrix}
0 & \L_\xi P_{\mu\nu} & 0 \\[2mm]
0 & \L_\xi {\Gamma^\rho}_{\mu\nu} + \d_\mu(\d_\nu\xi^\rho) & * \\[2mm]
0 & \L_\xi g_{\mu\nu} & 0 
\end{pmatrix} dx^\mu.
\end{align}
The entries are the correct infinitesimal transformations under active diffeomorphisms of the metric tensor, the Christoffel symbols and Schouten tensor. Entry $(2, 3)$ is of course redundant with entries $(1, 2)$ and entry $(3,2)$. 
 In the same way, 
 \begin{align}
 \label{expl_sig_Omega_0}
 \L_\xi \Omega_0&=L_\xi \Omega_0 + [\Omega_0, v_\xi] \notag\\
 			 &=L_\xi \begin{pmatrix} 0 & C & 0 \\ 0 & W & g\-\!\cdot\! C^T \\ 0 & 0 & 0\end{pmatrix} + \left[\begin{pmatrix} 0 & C & 0 \\ 0 & W & g\-\!\cdot\! C^T \\ 0 & 0 & 0\end{pmatrix}, \begin{pmatrix} 0 & 0 & 0\\  0 & \d\xi & 0 \\ 0 & 0& 0 \end{pmatrix} \right] \notag\\[3mm]
 			 &= \setlength{\arraycolsep}{7pt}\begin{pmatrix} 0  &  L_\xi C + C \cdot \d\xi & 0  \\[2mm] 0  &  L_\xi W + [W, \d\xi] & L_\xi(g\-\!\cdot\! C^T) - \d\xi \!\cdot\! (g\-\!\cdot\!C^T) \\[2mm] 0 & 0 & 0  \end{pmatrix} \notag\\[3mm]
 			 = \tfrac{1}{2} &\!\begin{pmatrix} 0  &  \xi^\alpha \d_\alpha C_{\nu, \mu\sigma}\! + C_{\nu, \alpha\sigma} \d_\mu\xi^\alpha \! +  C_{\nu, \mu\alpha} \d_\sigma\xi^\alpha \!+  C_{\alpha, \mu\sigma} \d_\nu\xi^\alpha \! & 0  \\[2mm] 0  & \xi^\alpha \d_\alpha {W^\rho}_{\nu, \mu\sigma}\! + {W^\rho}_{\nu, \alpha\sigma} \d_\mu\xi^\alpha \! +  {W^\rho}_{\nu, \mu\alpha} \d_\sigma\xi^\alpha \!+  {W^\rho}_{\alpha, \mu\sigma} \d_\nu\xi^\alpha \! -\d_\alpha \xi^\rho {W^\alpha}_{\nu, \mu\sigma} & * \\[2mm] 0 & 0 & 0  \end{pmatrix}\!dx^\mu\!\w\! dx^\sigma \notag\\[3mm]
=: \tfrac{1}{2} &\setlength{\arraycolsep}{7pt}\begin{pmatrix} 0 & \L_\xi C_{\nu, \mu\sigma} & 0 \\[2mm] 0 &  \L_\xi {W^\rho}_{\nu, \mu\sigma} & * \\[2mm] 0 & 0 & 0 \end{pmatrix} dx^\mu\w dx^\sigma. 
 \end{align}
The entries are the correct infinitesimal transformations under active diffeomorphisms of the Cotton and Weyl tensors.  Entry $(2, 3)$ is redundant with entry $(1, 2)$,  and entry $(3,2)$ of \eqref{expl_sig_varpi_0}. 
\bigskip

The computation in the more general (non-normal) case is just as easy
to perform. Beside the non zero terms in \eqref{expl_sig_Omega_0}, it
gives as entry $(2, 1)$ (and $(3,2)$) the Lie derivative of the
torsion tensor, $\tfrac{1}{2}\!\left(\L_\xi {T^\rho}_{\mu\sigma}\right)\! dx^\mu\w  dx^\sigma$, as in \eqref{R_and_T}, and as entry $(1, 1)$ (and $(3, 3)$) the transformation of the trace $f=\tfrac{1}{2}f_{\mu\nu} dx^\mu \w dx^\sigma= \tfrac{1}{2}P_{[\mu\nu]} dx^\mu \w dx^\sigma$, which is redundant with entry $(1, 2)$ in \eqref{expl_sig_varpi_0}.

\section{Conclusion} 
\label{Conclusion} 

In this paper we have briefly summed up the heuristic method that allows to built a BRST algebra describing a mixed gauge + diffeomorphism symmetry, $\text{BRST}^\xi$. A process we referred to as ``shifting'', according to~\cite{Bertlmann}, of the initial pure gauge algebra $\text{BRST}$. 
\smallskip

Then we have shown that the shifting method is compatible with the
dressing field approach, the latter was already shown to fit the BRST framework in~\cite{FLM15}. Two possibilities were in order: either dressing first then shifting and finding $\h{\text{BRST}}^\xi$, or shifting
first then dressing and finding $\h{\text{BRST}^\xi}$. We highlighted the
necessary and sufficient condition the dressing field has to satisfy
so as to warrant the commutation of these two fields redefinitions. In the case where this condition is not fulfilled, the treated
examples indicate that the correct algebra is always $\h{\text{BRST}^\xi}$ since it is the one taking into account the possible tensorial nature of the dressing field. 
\smallskip

Two instances of Cartan geometries illustrate how the general scheme, which consists in shifting first and then dressing, allows to recover efficiently, using a compact matrix formalism,
relevant results about mixed (gauge + diffeomorphisms) BRST algebras of gravitational theories. 

The first example was concerned with General Relativity. 
We exhibited the mixed Lorentz + $\Diff(\M)$ BRST algebra satisfied by the Cartan connection and its curvature viz.~\cite{Langouche-Schucker-Stora,Stora:2005tp}. 
Applying the dressing field method to this BRST algebra completely gets rid of the Lorentz (gauge) sector, leaving only $\Diff(\M)$ as the residual symmetry to be described, that is the gauge group for GR. This is consistent with the results of \cite{GaugeInv}. 

The second example dealt with the second-order conformal structure, as a Cartan geometry. It ought to be suited for conformal (Weyl) gravity. The shifted BRST algebra describes the $H$ + $\Diff(\M)$ (see \eqref{HKK}) transformations of the conformal Cartan connection $\varpi$, and its curvature $\Omega$. In that case, two dressing fields were needed. 
After the first dressing we ended up with a BRST algebra describing
the $CO(1, m-1)+\Diff(\M)$ symmetry of the dressed Cartan connection,
$\varpi_1$, and its curvature $\Omega_1$. This stage illustrated the
commutation of the two operations.
Then, the second dressing finally provided the final BRST algebra
describing the residual Weyl + $\Diff(\M)$ symmetry of the normal
conformal Cartan connection in its Riemannian parametrization,
$\varpi_0$, and its curvature $\Omega_0$. In this parametrization
$\varpi_0$ and $\Omega_0$ contain respectively, on the one hand, the
metric and Schouten tensors and the Christoffel symbols, and, on the
other hand, the 
Cotton and Weyl tensors. The infinitesimal Weyl and $\Diff(\M)$ transformations of these objects are then easily found from our final mixed BRST algebra whose central object is the ghost ${v'}_0$~\eqref{v'_0}. The algebraic connection $\varpi_0+{v'}_0$ provides a geometric framework for the cohomological results given in~\cite{Boulanger1}. The final mixed BRST algebra we have derived is relevant for Weyl gravity and might be useful in the characterization of a mixed Weyl+$\Diff(\M)$ anomaly. 
\medskip

The scheme presented in the paper is robust, handy and yields relevant results. 
However, it relies on the intertwining ansatz~\eqref{sigma} for which one sees little mathematical basis apart from the soundness of the outcoming results.  This ansatz deserves to be better understood, 
for instance, thanks to a way to derive the mixed BRST algebra (of the type given in~\cite{Langouche-Schucker-Stora,Stora:2005tp} which requires the use of a background connection) from a well-grounded geometrical framework. This issue is under study.


\section*{Acknowledgments}

The present work has been partly stimulated by
discussions and correspondence with the late R.~Stora on some issues on the mixed BRST algebra. 
One of us (JF) wishes to thank the Riemann Center for their warm hospitality and for financial support. We would like to thanks J.~Stasheff for some comments on a former version of the manuscript.

This work has been carried out in the framework of the Labex ARCHIMEDE (Grant No. ANR-11-LABX-0033)
and of the A*MIDEX project (Grant No. ANR-11-IDEX-0001-02), funded by the “Investissements d’Avenir”
French Government program managed by the French National Research Agency (ANR).


\bibliographystyle{unsrtnat}
\bibliography{Biblio}

\begin{thebibliography}{34}
\providecommand{\natexlab}[1]{#1}
\providecommand{\url}[1]{\texttt{#1}}
\expandafter\ifx\csname urlstyle\endcsname\relax
  \providecommand{\doi}[1]{doi: #1}\else
  \providecommand{\doi}{doi: \begingroup \urlstyle{rm}\Url}\fi

\bibitem[Boulanger(2005)]{Boulanger1}
N.~Boulanger.
\newblock {A Weyl-covariant tensor calculus}.
\newblock \emph{J. Math. Phys.}, 46:\penalty0 053508, 2005.

\bibitem[Yang(2005)]{YangSelPap}
C.N. Yang.
\newblock \emph{{Selected Papers (1945-1980), with Commentary}}.
\newblock World Scientific Publishing Company, 2005.

\bibitem[Stora(1984)]{Sto84}
R.~Stora.
\newblock {Algebraic structure and topological origin of chiral anomalies}.
\newblock In G.~{'t Hooft} and al., editors, \emph{{Progress in Gauge Field
  Theory, Carg{\`e}se 1983}}, {NATO ASI Ser.B, Vol.115}. Plenum Press, 1984.

\bibitem[Baulieu and Thierry-Mieg(1984)]{Baulieu:1984pf}
L.~Baulieu and J.~Thierry-Mieg.
\newblock {Algebraic Structure of Quantum Gravity and the Classification of the
  Gravitational Anomalies}.
\newblock \emph{Phys. Lett.}, B145:\penalty0 53, 1984.
\newblock \doi{10.1016/0370-2693(84)90946-8}.

\bibitem[Baulieu and Bellon(1986)]{Baulieu:1985md}
L.~Baulieu and M.~Bellon.
\newblock {$p$-Forms and Supergravity: Gauge Symmetries in Curved Space}.
\newblock \emph{Nucl. Phys.}, B266:\penalty0 75, 1986.
\newblock \doi{10.1016/0550-3213(86)90178-1}.

\bibitem[Langouche et~al.(1984)Langouche, Sch{\"u}cker, and
  Stora]{Langouche-Schucker-Stora}
F.~Langouche, T.~Sch{\"u}cker, and R.~Stora.
\newblock {Gravitational Anomalies of the Adler-Bardeen Type}.
\newblock \emph{Phys. Lett.}, B145:\penalty0 342--346, 1984.

\bibitem[Bertlmann(1996)]{Bertlmann}
R.~A. Bertlmann.
\newblock \emph{{Anomalies In Quantum Field Theory}}, volume~91 of
  \emph{{International Series of Monographs on Physics}}.
\newblock Oxford University Press, 1996.

\bibitem[Fournel et~al.(2014)Fournel, Fran\c{c}ois, Lazzarini, and
  Masson]{GaugeInv}
C.~Fournel, J.~Fran\c{c}ois, S.~Lazzarini, and T.~Masson.
\newblock {Gauge invariant composite fields out of connections, with examples}.
\newblock \emph{Int. J. Geom. Meth. Mod. Phys.}, 11\penalty0 (1):\penalty0
  1450016, 2014.
\newblock \doi{10.1142/S0219887814500169}.

\bibitem[Fran\c{c}ois et~al.(2015{\natexlab{a}})Fran\c{c}ois, Lazzarini, and
  Masson]{NucleonSpin}
J.~Fran\c{c}ois, S.~Lazzarini, and T.~Masson.
\newblock {Nucleon spin decomposition and differential geometry}.
\newblock \emph{Phys. Rev.}, D91:\penalty0 045014, 2015{\natexlab{a}}.

\bibitem[Fran\c{c}ois et~al.(2015{\natexlab{b}})Fran\c{c}ois, Lazzarini, and
  Masson]{FLM15}
J.~Fran\c{c}ois, S.~Lazzarini, and T.~Masson.
\newblock {Residual Weyl symmetry out of conformal geometry and its BRST
  structure}.
\newblock \emph{JHEP}, 09:\penalty0 195, 2015{\natexlab{b}}.
\newblock doi:10.1007/JHEP09(2015)195,.

\bibitem[Fran\c{c}ois(2014)]{JF_PhD}
J.~Fran\c{c}ois.
\newblock \emph{{Reductions of gauge symmetries: a new geometrical approach}}.
\newblock PhD thesis, Aix-Marseille University, September 2014.

\bibitem[Ne'eman et~al.(1978)Ne'eman, Regge, and Thierry-Mieg]{Ne'eman:1978gg}
Y.~Ne'eman, T.~Regge, and J.~Thierry-Mieg.
\newblock {Ghost-Fields, BRS and Extended Supergravity as Applications of Gauge
  Geometry}.
\newblock In R.~Ruffini and al., editors, \emph{{Matter particles}}, pages
  301--303, 1978.

\bibitem[Baulieu and Thierry-Mieg(1982)]{Baulieu:1981sb}
L.~Baulieu and J.~Thierry-Mieg.
\newblock {The Principle of BRS Symmetry: An Alternative Approach to Yang-Mills
  Theories}.
\newblock \emph{Nucl. Phys.}, B197:\penalty0 477, 1982.
\newblock \doi{10.1016/0550-3213(82)90454-0}.

\bibitem[Stora(2006)]{Stora:2005tp}
R.~Stora.
\newblock {The Wess Zumino consistency condition: A Paradigm in renormalized
  perturbation theory}.
\newblock \emph{Fortsch. Phys.}, 54:\penalty0 175--182, 2006.
\newblock \doi{10.1002/prop.200510266}.

\bibitem[Bandelloni(1988)]{Ban88}
G.~Bandelloni.
\newblock {Diffeomorphism cohomology in Quantum Field Theory models}.
\newblock \emph{Phys. Rev.}, {\bf D38}:\penalty0 1156, 1988.

\bibitem[Barnich et~al.(2000)Barnich, Brandt, and Henneaux]{Barnich:2000zw}
G.~Barnich, F.~Brandt, and M.~Henneaux.
\newblock {Local BRST cohomology in gauge theories}.
\newblock \emph{Phys. Rept.}, 338:\penalty0 439--569, 2000.
\newblock \doi{10.1016/S0370-1573(00)00049-1}.

\bibitem[Becchi et~al.(1976)Becchi, Rouet, and Stora]{BRS-76}
C.~Becchi, A.~Rouet, and R.~Stora.
\newblock {Renormalization of Gauge Theories}.
\newblock \emph{Ann. Phys.}, 98:\penalty0 287--321, 1976.

\bibitem[Dubois-Violette(1986)]{DbV86}
M.~Dubois-Violette.
\newblock {The Weil-BRS algebra of a Lie algebra and the anomalous terms in
  gauge theory}.
\newblock \emph{J. Geom. Phys.}, {\bf 3}:\penalty0 525--565, 1986.

\bibitem[Stora()]{Stora_private_comm}
R.~Stora.
\newblock {Private communication.}

\bibitem[Milnor(1984)]{Mil84}
J.~Milnor.
\newblock {Remarks on infinite-dimensional {L}ie groups}.
\newblock In B.S. DeWitt and R.~Stora, editors, \emph{{Relativity, groups and
  topology {II}}}, {Les Houches}, pages 1009--1057. Elsevier Science
  Publishers, 1984.
\newblock Session XL, 1983.

\bibitem[Masson and Wallet(2011)]{Masson-Wallet}
T.~Masson and J.~C. Wallet.
\newblock {A Remark on the Spontaneous Symmetry Breaking Mechanism in the
  Standard Model}.
\newblock arXiv:1001.1176, 2011.

\bibitem[Garajeu et~al.(1995)Garajeu, Grimm, and
  Lazzarini]{Garajeu-Grimm-Lazzarini}
D.~Garajeu, R.~Grimm, and S.~Lazzarini.
\newblock {W-gauge structures and their anomalies: An algebraic approach}.
\newblock \emph{J. Math. Phys.}, 36:\penalty0 7043--7072, 1995.

\bibitem[Lazzarini and Tidei(2008)]{Lazzarini-Tidei}
S.~Lazzarini and C.~Tidei.
\newblock {Polyakov soldering and second order frames : the role of the Cartan
  connection}.
\newblock \emph{Lett. Math. Phys.}, 85:\penalty0 27--37, 2008.

\bibitem[Blagojevi{\'c} and Hehl(2013)]{Blagojevic:2013xpa}
M.~Blagojevi{\'c} and F.W. Hehl, editors.
\newblock \emph{{Gauge Theories of Gravitation. A Reader with commentaries}}.
\newblock {Imperial College Press}. World Scientific, 2013.

\bibitem[Sharpe(1997)]{Sharpe}
R.W. Sharpe.
\newblock \emph{{Differential geometry: Cartan's generalization of Klein's
  erlangen program}}, volume 166 of \emph{{Graduate texts in mathematics}}.
\newblock Springer, New-York, Berlin, Heidelberg, 1997.

\bibitem[Kobayashi(1972)]{Kobayashi}
S.~Kobayashi.
\newblock \emph{{Transformation groups in differential geometry}}.
\newblock {Classics in Mathematics, vol. {\bf 70}}. Spinger-Verlag, Berlin,
  1972.

\bibitem[Bonora et~al.(1984)Bonora, Pasti, and Tonin]{Bonora:1984pz}
L.~Bonora, P.~Pasti, and M.~Tonin.
\newblock {Gravitational and Weyl Anomalies}.
\newblock \emph{Phys. Lett.}, B149:\penalty0 346, 1984.
\newblock \doi{10.1016/0370-2693(84)90421-0}.

\bibitem[Bonora et~al.(1986{\natexlab{a}})Bonora, Pasti, and
  Tonin]{Bonora:1984ic}
L.~Bonora, P.~Pasti, and M.~Tonin.
\newblock {The Anomaly Structure of Theories With External Gravity}.
\newblock \emph{J. Math. Phys.}, 27:\penalty0 2259, 1986{\natexlab{a}}.
\newblock \doi{10.1063/1.526998}.

\bibitem[Bonora et~al.(1986{\natexlab{b}})Bonora, Pasti, and
  Bregola]{Bonora-Pasti-Bregola:1986}
L.~Bonora, P.~Pasti, and M.~Bregola.
\newblock {Weyl cocycles}.
\newblock \emph{Class. Quant. Grav.}, 3:\penalty0 635--649, 1986{\natexlab{b}}.

\bibitem[Moritsch and Schweda(1994)]{Moritsch:1994hv}
O.~Moritsch and M.~Schweda.
\newblock {On the algebraic structure of gravity with torsion including Weyl
  symmetry}.
\newblock \emph{Helv. Phys. Acta}, 67:\penalty0 289--344, 1994.

\bibitem[Boulanger(2007{\natexlab{a}})]{Boulanger2}
N.~Boulanger.
\newblock {Algebraic Classification of Weyl Anomalies in Arbitrary Dimensions}.
\newblock \emph{Phys. Rev. Lett.}, 98:\penalty0 261302, 2007{\natexlab{a}}.

\bibitem[Boulanger(2007{\natexlab{b}})]{Boulanger:2007st}
N.~Boulanger.
\newblock {General solutions of the Wess-Zumino consistency condition for the
  Weyl anomalies}.
\newblock \emph{JHEP}, 0707:\penalty0 069, 2007{\natexlab{b}}.
\newblock \doi{10.1088/1126-6708/2007/07/069}.

\bibitem[Ogiue(1967)]{Ogiue}
K.~Ogiue.
\newblock {Theory of Conformal Connections}.
\newblock \emph{Kodai Math. Sem. Rep.}, 19:\penalty0 193--224, 1967.

\bibitem[Go et~al.(1974)Go, Kastrup, and Mayer]{Go:1971zz}
T.~H. Go, H.~A. Kastrup, and D.~H. Mayer.
\newblock {Properties of dilatations and conformal transformations in Minkowski
  space}.
\newblock \emph{Rept. Math. Phys.}, 6:\penalty0 395--430, 1974.
\newblock \doi{10.1016/S0034-4877(74)80006-6}.
\newblock doi:10.1016/S0034-4877(74)80006-6.

\end{thebibliography}
\addcontentsline{toc}{section}{References}

\end{document}